\documentclass[12pt]{article}
\usepackage{graphicx}
\usepackage[utf8]{inputenc}
\usepackage{mathtools}
\usepackage{multirow}
\usepackage{graphicx,psfrag,epsf,color}
\usepackage{amsmath,amssymb,amsfonts}
\usepackage{array}
\usepackage{cite}
\usepackage{float}
\usepackage{enumitem}
\usepackage{appendix}
\usepackage{bm}
\usepackage{ascii}
\usepackage{dsfont}
\usepackage{caption}
\usepackage{subcaption}
\usepackage[colorlinks]{hyperref}
\usepackage[dvipsnames]{xcolor}
\hypersetup{pageanchor=false,linkcolor=Maroon,citecolor=MidnightBlue,urlcolor=MidnightBlue}
\usepackage{array}

\renewcommand{\Re}{\textrm{Re}}

\renewcommand{\vec}[1]{\bm{#1}}
\newcommand{\cor}[2]{#1\langle #2 #1\rangle}
\newcommand{\p}{\partial}
\newcommand{\vp}{\varphi}

\newcommand{\der}{\textrm{d}}
\newcommand{\Lo}{\mathcal{O}}

\newcommand{\N}{\mathbb{N}}
\newcommand{\Z}{\mathbb{Z}}

\newcommand{\R}{\mathbb{R}}
\newcommand{\C}{\mathbb{C}}
\newcommand{\D}{\mathcal{D}}

\newcommand{\sG}{\mathcal{G}}
\newcommand{\secref}[1]{Sec.~\ref{#1}}
\newcommand{\figref}[1]{Fig.~\ref{#1}}
\newcommand{\appref}[1]{App.~\ref{#1}}
\newcommand{\tabref}[1]{Tab.~\ref{#1}}
\renewcommand{\log}{\ln}

\newcounter{MBQ}

\newcounter{ASQ}

\numberwithin{equation}{section}

\begin{document}
\allowdisplaybreaks

\begin{titlepage}

\begin{flushright}
{\small
TUM-HEP-1616/26\\
October 11, 2026\\
}
\end{flushright}

\vskip1cm
\begin{center}
{\Large \bf On the analytic continuation and asymptotic\\[0.0cm] 
behaviour of the two-point function in massless\\[0.15cm] 
$\phi^4$-theory in Euclidean de Sitter space}
\end{center}
  \vspace{0.5cm}
\begin{center}
{\sc Martin~Beneke$^{a}$ and Andrea~F.~Sanfilippo$^{b}$} 
\\[6mm]
{\it ${}^a$Physik Department T31,\\
James-Franck-Stra\ss e~1,
Technische Universit\"at M\"unchen,\\
D--85748 Garching, Germany}\\[0.2cm]
{\it ${}^b$Departamento de F\'isica Te\'orica y del Cosmos, Universidad de Granada,}\\
{\it Campus de Fuentenueva, E–18071 Granada, Spain}
\end{center}
\vskip1cm

\begin{abstract}
\noindent
For the two-point function of the massless, minimally-coupled, real scalar field in Euclidean de Sitter space with $\lambda\phi^4$-interaction, we compute the leading-logarithmic terms in the invariant distance of the two-point function to $\Lo(\lambda^2)$, which counts as the fifth subleading order (N$^5$LO) in the $\sqrt{\lambda}$-expansion. We show that the analytic continuation to Lorentzian signature coincides with the expansion of the analogous quantity computed in the stochastic formalism. We determine the large-order behaviour of the stochastic series expansion and prove its Borel summability, which suggests that the Euclidean theory can be used to extract the late-time behaviour of massless correlation functions in Lorentzian de Sitter space.

\end{abstract}

\end{titlepage}


\section{Introduction}

Correlation functions of minimally coupled, massless scalar fields in de Sitter (dS) spacetime suffer from well-known infrared (IR) divergences when computed in perturbation theory, irrespective of the number of spacetime dimensions $d$. This may be regarded as a reflection of the mathematical statement that there is no vacuum state of the free massless theory which respects all de Sitter isometries \cite{Allen:1985ux}. Physically, since the theory in question does not enjoy conformal invariance, gravitational particle production is expected to occur, and field modes with small physical momentum  $\vec k/(a(t)H)$ in Hubble units, the superhorizon modes, get copiously produced. Due to this effect, the amplitude of the free massless field is amplified without bound by the expanding spacetime, and its correlation functions diverge. Once an interaction term of some kind is included in the theory, it is plausible that it will eventually counteract the IR enhancement of the field and stabilise the theory. In that case, however, it may also be expected that the stable state of the interacting theory, which should then be identified as the true vacuum state, is of non-perturbative nature. 

This interesting and non-trivial phenomenon was first studied in the massless $\lambda\phi^4$-theory in dS through the formalism of stochastic inflation \cite{Starobinsky:1982ee,Starobinsky:1986fx}, which culminated in the seminal paper \cite{Starobinsky:1994bd}. The superhorizon modes of the full quantum field $\phi$ are described in terms of a classical, stochastic random variable $\bar\phi$, which is governed by a Langevin equation. The subhorizon part of the field no longer appears as a dynamical variable. However, its presence manifests itself as the source of Gaussian white noise in the Langevin equation. 
This approach allows for the non-perturbative computation of correlation functions of $\bar\phi$, and it uncovers the dynamical generation of the mass scale $\lambda^{1/4} H$ as the mechanism for the self-regularisation of the IR divergences observed in perturbation theory. The distinctive non-analytic dependence of this scale on the fourth root of the coupling, rather than on $\lambda$ itself, makes its non-perturbative origin manifest. The stochastic treatment was shown  in \cite{Garbrecht:2014dca} to be equivalent to the leading IR behaviour of the superhorizon degrees of freedom in a quantum-field-theoretic description. The full set of next-to-leading corrections was obtained in \cite{Gorbenko:2019rza,Mirbabayi:2020vyt}, and some of the next-to-next-to-leading corrections were computed in \cite{Cohen:2021fzf,Beneke:2026ksj}.

Complementary insight can be obtained from the Euclidean version of the theory. Euclidean de Sitter (EdS) space is equivalent to the $d$-dimensional sphere with radius $1/H$. Since the sphere is compact, quantum fields can be expanded in terms of discrete modes labeled by their (generalised) angular momentum $\vec L$. It was realised in \cite{Rajaraman:2010xd} that in this setting all IR divergences of the free massless scalar theory can be traced back to the constant mode with vanishing angular momentum, the zero mode $\phi_0$. It was shown that, once a self-interaction is turned on, the exact treatment of the zero mode and the perturbative treatment of all other modes leads to IR-finite correlation functions of the massless field. As a consequence of the non-perturbative treatment of $\phi_0$, the square root of the quartic coupling 
emerges as the natural expansion parameter of the theory. The subsequent work \cite{Beneke:2012kn} systematised the 
$\sqrt{\lambda}$-expansion of Euclidean correlation functions in massless $\lambda\phi^4$-theory, providing the rules to compute IR-finite and manifestly rotationally-invariant correlation functions on the sphere.

This construction relies crucially on the discreteness of the mode expansion of the field. It therefore leaves the question of the relevance of the Euclidean results to physical dS space open. In \cite{Hollands:2011we} 
it was suggested to analytically continue the correlators computed on the sphere to Lorentzian signature by extending the allowed range of values of the kinematic invariants. This procedure has been proven to yield the desired physical quantities in the case of massive scalar fields in \cite{Higuchi:2010xt}. However, in the massless case it is not clear what the interpretation of the resulting quantity would be, and what it should be compared to in the Lorentzian setting. Furthermore, any fixed-order Euclidean result for a correlation function will typically contain logarithms of the relevant kinematic invariants, such as the de Sitter-invariant distance $Z_{xy}$ in the case of the two-point function $\cor{}{\phi(x)\phi(y)}$. These logarithms stay bounded on the sphere, but can grow arbitrarily large after analytic continuation. This is an avatar of the well-known problem of secular growth of massless correlation functions. In particular, in inflationary physics one is usually interested in equal-time correlation functions evaluated at late times, formally $t\rightarrow\infty$, which in the two-point function corresponds to letting $|Z_{xy}|\rightarrow\infty$. To obtain something physically meaningful after analytic continuation, the resummation of such large logarithmic terms is mandatory. 

Refs.~\cite{LopezNacir:2016gzi,LopezNacir:2018xto,LopezNacir:2019ord} addressed this point by generalising the formalism of \cite{Beneke:2012kn} to the massless $O(N)$ vector model. The two-point function of the $N$-component field $\phi_a$ was computed at leading \cite{LopezNacir:2016gzi} and next-to-leading \cite{LopezNacir:2018xto,LopezNacir:2019ord} order in the $1/N$-expansion and to all orders in $\sqrt{\lambda}$, thereby achieving the necessary resummation of logarithmic terms and making it suitable for analytic continuation. The large-$Z_{xy}$ behaviour of the analytically continued two-point function was studied, and up to the considered order it was found to behave like a sum of two free massive propagators with two different dynamically generated masses of order $\lambda^{1/4} H$. This result is consistent with the Lorentzian large-$N$ results of \cite{Gorbenko:2019rza} and \cite{Gautier:2013aoa,Gautier:2015pca}. Furthermore, the found dynamical masses coincide with the relaxation eigenvalues obtained from the large-$N$ version of stochastic inflation 
\cite{Moreau:2020gib}. These studies provide evidence for the equivalence of the resummed Euclidean and Lorentzian results after analytic continuation of the former in the case of the $O(N)$ model at large $N$.

This paper explores to which extent this equivalence carries over to the single-field case. We study the analytic continuation of the Euclidean two-point function in massless $\lambda\phi^4$-theory, in the limit of large invariant distance $Z_{xy}$. We determine the leading terms in the $|Z_{xy}|\rightarrow\infty$ limit up to $\Lo(\lambda^2)$, i.e.~the fifth subleading order (N$^5$LO) in the $\sqrt{\lambda}$-expansion. We then show that the result for the Taylor expansion of the equilibrium two-point function of the stochastic long-wavelength field $\bar\phi$ is reproduced up to this order. By providing the asymptotic behaviour numerically and a proof of Borel summability of the series representation for the stochastic two-point function, we then argue that the late-time behaviour of the full Lorentzian two-point function is encoded in the analytic continuation of the Euclidean result.

\section{Euclidean and stochastic two-point functions}

We begin by defining the two quantities to be studied and compared, which are the Euclidean two-point function $\cor{}{\phi(x)\phi(y)}$ and the stochastic equilibrium two-point function $\cor{}{\bar\phi(t_x,\vec x)\bar\phi(t_y,\vec y)}$. We briefly recall relevant aspects of the Euclidean and stochastic formalisms, allowing  for any spacetime dimension $d>1$,  since this entails no further complications compared to $d=4$.

\subsection{Massless two-point function on the sphere}
\label{sec::EdS}

The theory of interest is defined by the Euclidean action
\begin{equation}
S_E=\int\der^dx\;\sqrt{g(x)}\bigg[\,\frac{1}{2}g^{\mu\nu}(x)\p_{\mu}\phi(x)\p_{\nu}\phi(x)+\frac{\lambda}{4!}\phi^4(x)\bigg]\,,
\end{equation}
where $g_{\mu\nu}(x)$ is the metric of the $d$-dimensional sphere with radius $1/H$. Following~\cite{Beneke:2012kn}, the field is split as
\begin{equation}
\phi(x)=\phi_0+\hat\phi(x)\,,
\label{eq::fieldsplit}
\end{equation}
where the zero-mode part $\phi_0$ is a real-valued constant. This leads to the action
\begin{equation}
S_E=S_{\textrm{free}}+S_{\textrm{int}}\,,
\end{equation}
where
\begin{flalign}
S_{\textrm{free}}&\equiv\frac{V_d\lambda}{4!}\phi^4_0+\frac{1}{2}\int\der^dx\;\sqrt{g(x)}\,g^{\mu\nu}(x)\p_{\mu}\hat\phi(x)\p_{\nu}\hat\phi(x)\,,\label{eq::Sfree}\\
S_{\textrm{int}}&\equiv\int\der^dx\;\sqrt{g(x)}\,\bigg[\,\frac{\lambda}{2}\phi^2_0\hat\phi(x)^2+\frac{\lambda}{3!}\phi_0\hat\phi(x)^3+\frac{\lambda}{4!}\hat\phi(x)^4\bigg]\,,\label{eq::Sint}
\end{flalign}
and 
\begin{equation}
V_d\equiv\frac{2\pi^{\frac{d+1}{2}}}{H^d\Gamma(\frac{d+1}{2})}
\label{eq::EdSvolume}
\end{equation}
is the volume of EdS. The $\phi_0^3\hat{\phi}(x)$ interaction can be dropped due to \eqref{eq::Yvanish}.
Notice that the pure zero-mode interaction term proportional to $\lambda\phi^4_0$ has been included in the ``free" action \eqref{eq::Sfree}. This implies the power-counting $\phi_0\sim\lambda^{-\frac{1}{4}}$ for the zero mode, while the non-zero-mode field counts as $\hat\phi\sim\lambda^0$. As a consequence, the three terms in \eqref{eq::Sint} count as $\sqrt{\lambda}$, $\lambda^{\frac{3}{4}}$, and $\lambda$, respectively, and can therefore be treated as small perturbations.

Correlation functions on the sphere are defined by the Euclidean path integral
\begin{equation}
\cor{}{\phi(x_1)...\phi(x_n)}=\frac{\int\D[\phi]e^{-S_E}\phi(x_1)...\phi(x_n)}{\int\D[\phi]e^{-S_E}}\,.
\label{eq::EucPathInt}
\end{equation}
To compute massless Euclidean correlation functions in the $\sqrt{\lambda}$-expansion we implement the field split \eqref{eq::fieldsplit} for all fields appearing in the correlation function \eqref{eq::EucPathInt}, and expand the path-integral measure and weight as
\begin{equation}
\int\D[\phi]e^{-S_E}=\sum_{n=0}^{\infty}\frac{(-1)^n}{n!}\int_{-\infty}^{\infty}\der\phi_0\int\D[\hat\phi]\,e^{-S_{\textrm{free}}}(S_{\textrm{int}})^n\,.
\label{eq::pertexp}
\end{equation}
The zero-mode part of the path-integral weight is inherently non-Gaussian, so zero-mode correlation functions cannot be given a diagrammatic representation. To proceed, all factors of $\phi_0$ from $(S_{\textrm{int}})^n$  and from the external fields appearing in the integrand of \eqref{eq::EucPathInt} are collected, and the resulting pure zero-mode correlation function is computed exactly with the formula
\begin{equation}
\cor{}{\phi^{2n}_0}_0\equiv\bigg[\int_{-\infty}^{\infty}\der\phi_0\;e^{-\frac{V_d\lambda}{4!}\phi^4_0}\bigg]^{-1}\int_{-\infty}^{\infty}\der\phi_0\;e^{-\frac{V_d\lambda}{4!}\phi^4_0}\phi^{2n}_0=\bigg(\frac{4!}{\lambda V_d}\bigg)^{\!\frac{n}{2}}\,\frac{\Gamma(\frac{2n+1}{4})}{\Gamma(\frac{1}{4})}\,.
\label{eq::0cor}
\end{equation}
Correlation functions of odd powers of $\phi_0$ vanish due to the $\Z_2$-symmetry under $\phi\rightarrow-\phi$ of the action. On the other hand, the non-zero mode part of $S_{\rm free}$ is Gaussian, so the integration over $\hat\phi$ in every term in the expansion of \eqref{eq::EucPathInt} can be represented diagrammatically in the usual way by lines connecting all fields in the integrand pairwise to each other. These lines correspond to the free non-zero mode propagator
\begin{equation}
\hat G(x,y)\equiv\cor{}{\hat\phi(x)\hat\phi(y)}_{|\,\Lo(\lambda^0)}\,,
\label{eq::EGHatd}
\end{equation}
which is given by \cite{Beneke:2012kn}
\begin{equation}
\hat G(x,y)= H^{d-2}\sum_{\vec L\neq0}\frac{Y_{\vec L}(x)Y^*_{\vec L}(y)}{L(L+d-1)}\,,
\label{eq::Gspec}
\end{equation} 
with $Y_{\vec L}(x)$ the $d$-dimensional hyperspherical harmonics, see \appref{app::hyperY} for our conventions and their basic properties. The sum evaluated in $d=4$ dimensions reads \cite{Allen:1986tt}
\begin{equation}
\hat G(x,y)=\frac{H^2}{8\pi^2}\bigg[\frac{1}{1-Z_{xy}}-\log\bigg(\frac{1-Z_{xy}}{2}\bigg)-\frac{7}{3}\bigg]\,.
\label{eq::EGHat4d}
\end{equation}
In this formula $Z_{xy}$, defined as
\begin{equation}
Z_{xy}\equiv\cos(\theta_x)\cos(\theta_y)+\sin(\theta_x)\sin(\theta_y)\,\vec x\cdot\vec y\,,
\end{equation} 
denotes the embedding distance in the ambient space $\R^{d+1}$ between the points $x$ and $y$, parametrised by the polar angles $\theta_{x,y}$ and the unit vectors $\vec x,\,\vec y$ on the sub-sphere $S^{d-1}$.
For points on the sphere, $Z_{xy}$ is restricted to lie in the compact interval $Z_{xy}\in[-1,1]$. In this set-up, we will study the interacting Euclidean two-point function,
\begin{equation}
G(x,y)\equiv\cor{}{\phi(x)\phi(y)} = \cor{}{\phi_0\phi_0}+\cor{}{\hat\phi(x)\hat\phi(y)}\,,
\end{equation}
which, in the present treatment, decomposes into a zero-mode and non-zero-mode part as shown after the second equality.
Due to the $d$-dimensional rotational invariance, the result can only depend on $Z_{xy}$.

\subsection{Stochastic equilibrium two-point function}
\label{sec::SI}

We wish to compare the analytic continuation of the Euclidean two-point function in $d$ dimensions to the two-point in-in correlation function of the massless scalar field computed directly in Lorentzian signature. To do so, we use the formalism of stochastic inflation \cite{Starobinsky:1994bd}. In the stochastic paradigm, the superhorizon modes of the full scalar field are described by the random variable $\bar\phi(t,\vec x)$, which satisfies the Langevin equation
\begin{equation}
\frac{\p}{\p t}\bar\phi(t,\vec x)=-\frac{1}{(d-1)H}V'(\bar\phi)+f(t,\vec x)\,,
\end{equation}
where $f(t,\vec x)$ is a Gaussian white noise source with local variance
\begin{equation}
\cor{}{f(t,\vec x)f(t',\vec x)}=\frac{2}{(d-1)H V_d}\,\delta(t-t')\,.
\end{equation}
Remarkably, $V_d$ is the Euclidean volume given by \eqref{eq::EdSvolume}. All one-point correlation functions $\cor{}{\bar\phi^n(t,\vec x)}$ are encoded in the one-point probability-distribution function (PDF) $\rho_1(t,\vp)$, defined by 
\begin{equation}
\cor{}{\bar\phi^n(t,\vec x)}\equiv\int_{-\infty}^{\infty}\der\vp\;\rho_1(t,\vp)\,\vp^n\,.
\label{eq::stoch1pt}
\end{equation}
It satisfies the Fokker-Planck (FP) equation
\begin{equation}
\frac{\p}{\p t}\rho_1(t,\vp)=\Gamma_{\vp}\,\rho_1(t,\vp)\,,
\label{eq::FP}
\end{equation}
with the FP operator
\begin{equation}
\Gamma_{\vp}=\frac{1}{(d-1)H}\bigg[\frac{\p}{\p\vp}V'(\vp)+\frac{1}{V_d}\frac{\p^2}{\p\vp^2}\bigg]\,.
\label{eq::ddimGamma}
\end{equation} 
Eq.~\eqref{eq::FP} is the central equation of the stochastic formalism, as all higher $n$-point PDFs can be shown to be determined by it \cite{Starobinsky:1994bd}. Its  general solution may be expanded as
\begin{equation}
\rho_1(t,\vp)=e^{-v(\vp)}\sum_{n=0}^{\infty}a_n\Phi_n(\vp)e^{-\Lambda_n(t-t_0)}\,,\quad v(\vp)\equiv\frac{1}{2}V_d \,V(\vp)\,,
\label{eq::gen1PDF}
\end{equation}
where the $\Phi_n(\vp)$ form the complete set of orthonormal eigenfunctions of the Schr\"odinger-like equation
\begin{equation}
\frac{1}{2}\bigg[-\frac{\p^2}{\p\vp^2}+[v'(\vp)]^2-v''(\vp)\bigg]\Phi_n(\vp)=\frac{\Lambda_n(d-1)HV_d}{2}\,\Phi_n(\vp)\,.
\label{eq::Schrodi}
\end{equation}
The $\Lambda_n$ are called relaxation eigenvalues. They are non-negative, non-degenerate, and increasing in magnitude with $n$. The coefficients $a_n$ in \eqref{eq::gen1PDF} encode the value of $\rho_1(t_0,\vp)$ at some initial time $t_0$, which must be specified. Assuming that the PDF is normalisable, the lowest eigenvalue is $\Lambda_0=0$, and as $t\rightarrow\infty$ only the $n=0$ summand in \eqref{eq::gen1PDF} survives. This term defines the time-independent equilibrium solution $\rho_{1\textrm{eq}}(\vp)$, for which the left-hand side of \eqref{eq::FP} vanishes. Integrating $\Gamma_\vp \,\rho_{1\textrm{eq}}(\vp)=0$ and imposing the normalisation $\int_{-\infty}^{\infty}\der\vp\;\rho_{1\textrm{eq}}(\vp)=1$, one finds that for $V(\vp)=\frac{\lambda}{4!}\vp^4$ it is given by
\begin{equation}
\rho_{1\textrm{eq}}(\vp)=\frac{2}{\Gamma(\frac{1}{4})}\bigg(\frac{V_d\lambda}{4!}\bigg)^{\!\frac{1}{4}}\exp\bigg(-\frac{\lambda V_d}{4!}\vp^4\bigg)\,.
\label{eq::eqPDF}
\end{equation}
It was pointed out in \cite{Beneke:2012kn} that the equilibrium one-point functions $\cor{}{\bar\phi^{2n}}$ computed using \eqref{eq::eqPDF} agree with the values \eqref{eq::0cor} of the pure zero-mode correlation functions $\cor{}{\phi^{2n}_0}_0$ for any $d$. This observation may be regarded as a first glimpse of the connection between the Euclidean and stochastic results. The PDF $\rho_{1\textrm{eq}}(\vp)$ is also the only solution of \eqref{eq::FP} which yields a de Sitter-invariant (constant) result for the one-point functions \eqref{eq::stoch1pt}. As such, this is the solution of the FP equation that should be used to compute any $\cor{}{\bar\phi^n(t,\vec x)}$. Notice, however, that this does not imply that the $n\geq 1$ terms in \eqref{eq::gen1PDF} are unphysical, as \eqref{eq::FP} determines \textit{all} stochastic $n$-point correlation functions \cite{Starobinsky:1994bd}.

We are now in the position to define the counterpart to the Euclidean two-point function, which is the stochastic {\em equilibrium} two-point function
\begin{equation}
\cor{}{\bar\phi(t_x,\vec x)\bar\phi(t_y,\vec y)}\,.
\label{eq::stochGdef}
\end{equation}
In the following, the angle brackets will always refer to the stochastic average 
with the equilibrium distribution $\rho_{\rm 1eq}(\varphi)$ \eqref{eq::eqPDF}. 
Its value at coincident points $(t_x,\vec x)=(t_y,\vec y)$ is \begin{equation}
\cor{}{\bar\phi^2(t_x,\vec x)}\equiv\cor{}{\bar\phi^2}\,.
\label{eq::coincidentG}
\end{equation}
In \cite{Starobinsky:1994bd} it is shown that for both large time-like ($Z_{xy}\gg1$) and space-like ($Z_{xy}<0$, $|Z_{xy}|\gg1$) separation, which are the regimes in which the present formalism applies, this quantity can be written in the manifestly de Sitter-invariant form
\begin{equation}
\cor{}{\bar\phi(t_x,\vec x)\bar\phi(t_y,\vec y)}=\sG\Big(H^{-1}\log|2Z_{xy}-1|\Big)\,.
\label{eq::dSinv2pt}
\end{equation}
The right-hand side is expressed through the autocorrelation function
\begin{equation}
\sG(z)\equiv\sum_{n=0}^{\infty}b_ne^{-\Lambda_nz}\,.
\label{eq:curlyG}
\end{equation}
The $\Lambda_n$ are the relaxation eigenvalues introduced in \eqref{eq::Schrodi}, and the coefficients $b_n$ are given by 
\cite{Starobinsky:1994bd,Markkanen:2019kpv}
\begin{equation}
b_n\equiv\bigg[\int_{-\infty}^{\infty}\der\vp\;\vp\,\Phi_0(\vp)\Phi_n(\vp)\bigg]^2\,,
\label{eq::bn}
\end{equation}
with $\Phi_n(\vp)$ the eigenfunctions introduced in \eqref{eq::Schrodi}. The coincident limit $z=0$ is fixed by \eqref{eq::coincidentG} and translates into the constraint
\begin{equation}
\sum_{n=0}^{\infty}b_n=\cor{}{\bar\phi^2}\,,
\label{eq::bnconstraint}
\end{equation}
which  follows from \eqref{eq::dSinv2pt} and \eqref{eq:curlyG}. 

The stochastic formalism also allows one to compute $\sG$ 
in the form of a power-series expansion for small argument:
\begin{equation}
\sG(z)=\sum_{n=0}^{\infty}\frac{1}{n!}\frac{\der^n\sG(z)}{\der z^n}\bigg|_{z=0}z^n\,.
\label{eq::gexp}
\end{equation}
The Taylor coefficients may be computed analytically by means of the formula \cite{Starobinsky:1994bd}
\begin{equation}
\frac{\der^n\sG(z)}{\der z^n}\bigg|_{z=0}=\int_{-\infty}^{\infty}\der\vp\,\vp\,[(\Gamma_{\vp})^n\rho_{1\textrm{eq}}(\vp)\vp]\,,
\label{eq::taylorcoeff}
\end{equation}
with the FP operator $\Gamma_{\vp}$ defined in equation \eqref{eq::ddimGamma}.  Alternatively, from \eqref{eq:curlyG} they may be computed as 
\begin{equation}
\frac{\der^n\sG(z)}{\der z^n}\bigg|_{z=0}=\sum_{m=0}^{\infty}b_m(-\Lambda_m)^n\,.
\label{eq::taylorcoeff2}
\end{equation}


\section{Euclidean two-point function: leading logarithms up to $\Lo(\lambda^2)$}
\label{sec::Euc2pt}

We apply the Euclidean formalism discussed in \secref{sec::EdS} to the computation of the connected Green function
\begin{equation}
G(x,y)=\cor{}{\phi_0\phi_0}+\cor{}{\hat\phi(x)\hat\phi(y)}\,.
\end{equation}
Due to rotational invariance on the sphere, which becomes dS invariance after analytic continuation, the two-point function must be a function of $Z_{xy}$ alone. Since in the following we only consider two-point functions evaluated at the points $x$ and $y$, from now on we abbreviate $Z_{xy}=Z$. We wish to extract the leading contributions in the limit $|Z|\rightarrow\infty$, which are of the form $(\sqrt{\lambda})^n \ln^{n+1} |Z|$ at order 
$(\sqrt{\lambda})^n$. We will refer to this series of terms as the ``leading logarithms''. In this approximation, $G$ is represented by the series
\begin{equation}
G(x,y)_{|\,\textrm{LL}} = \sum_{n=-1}^{\infty} g_n \,(\sqrt{\lambda})^n \ln^{n+1} |Z|\,.
\label{eq::GLLseries}
\end{equation}
We develop a method suited to extract these terms from the relevant diagrams, which simplifies the computations considerably, uncovers the insensitivity of the leading logarithms to ultraviolet (UV) divergences, and allows us to compute them up to $\Lo(\lambda^2)$, which is done here for the first time.

\subsection{Identification of the relevant diagrams}
\label{sec::Ediaglist}

Before identifying the relevant diagrams for the computation of the leading logarithms of $G(x,y)$ up to $\Lo(\lambda^2)$, a comment about vacuum diagrams 
in the Euclidean formalism is in order. The non-Gaussianity of the zero-mode path integral implies that zero-mode correlation functions do not satisfy Wick's theorem, e.g. $\cor{}{\phi^4_0}_0\neq 3\cor{}{\phi^2_0}^2$. This leads to an incomplete cancellation of vacuum-diagram contributions between the numerator and denominator of \eqref{eq::EucPathInt}, and, in general, this must be taken into account. However, after renormalising the theory, these diagrams effectively constitute subleading (in $\sqrt{\lambda}$) corrections to the constant multiplying any term in the numerator of \eqref{eq::EucPathInt} and therefore do not contribute leading logarithms. Therefore, for the present purpose, they can be neglected.

\begin{figure}[t]
\centering
\begin{subfigure}{0.5\textwidth}
\centering
\includegraphics[width=0.5\textwidth]{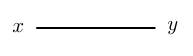}
\caption{}
\end{subfigure}%
\begin{subfigure}{0.5\textwidth}
\centering
\includegraphics[width=0.8\textwidth]{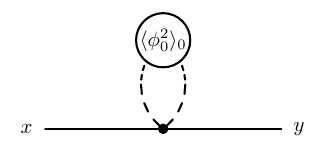}
\caption{}
\end{subfigure}
\caption{Free non-zero mode propagator (a) and $\Lo(\sqrt{\lambda})$ chain diagram (b).}
\label{fig::Ediags1}
\end{figure}

The leading contribution to $G(x,y)$ is the constant zero-mode correlation function $\cor{}{\phi_0\phi_0}$. At leading order in $\sqrt{\lambda}$, which is $\Lo(\lambda^{-\frac{1}{2}})$, it reduces to
\begin{equation}
\cor{}{\phi^2_0}_0=\sqrt{\frac{4!}{\lambda V_d}}\,\frac{\Gamma(\frac{3}{4})}{\Gamma(\frac{1}{4})}\,,
\label{eq::phi02}
\end{equation} 
where we used \eqref{eq::0cor}. Corrections to $\cor{}{\phi_0\phi_0}$ from non-zero mode loop diagrams are all suppressed by powers of $\sqrt{\lambda}$ and do not contain any powers of $\log|Z|$. Since we are only interested in the leading-logarithmic terms of the two-point function, we can ignore them. All of the relevant higher-order contributions will thus come from the non-zero-mode part $\cor{}{\hat\phi(x)\hat\phi(y)}$, to which we now turn. 

At $\Lo(\lambda^0)$ the non-zero mode two-point function is given by the free propagator \eqref{eq::EGHatd}. We draw it as a solid line connecting the points $x$ and $y$, as shown in \figref{fig::Ediags1} (a). At $\Lo(\sqrt{\lambda})$ we have the first non-trivial diagram, involving the insertion of the $\phi^2_0\hat\phi^2(x)$ vertex, shown in \figref{fig::Ediags1} (b). It leads to the expression 
\begin{equation}
-\frac{\lambda}{2}\cor{}{\phi^2_0}_0\int_z\hat G(x,z)\hat G(z,y)\,,
\end{equation}
where we introduced the shorthand notation 
\begin{equation}
\int_z\equiv\int\der^dz\,\sqrt{g(z)}\,.
\end{equation}
Following \cite{LopezNacir:2016gzi}, in the above diagram and in all the following diagrams, we represent the procedure of collecting all the factors of $\phi_0$ appearing in a given diagram, and then evaluating the resulting $\phi_0$-correlation functions using \eqref{eq::0cor}, pictorially by dashed lines leading from all vertices involving $\phi_0$ to a circle containing the resulting $\phi_0$-correlator. 

At each order in the expansion of $G(x,y)$ in $\sqrt{\lambda}$, we will always have one diagram involving only insertions of the $\phi^2_0\hat\phi^2(x)$ vertex. We will refer to this type of diagrams as ``chain diagrams", and we will call a diagram involving $n$ insertions of the $\phi^2_0\hat\phi^2(x)$ vertex an ``$n$-chain". Accordingly, the diagram in \figref{fig::Ediags1} (b) is a 1-chain, and $\hat G(x,y)$ may be regarded as a 0-chain. Since the structure of the integrals which result from the chain diagrams is always the same, it proves convenient to define the class of chain integrals
\begin{flalign}
I_{0\textrm{-chain}}(Z)\equiv&\;\hat G(x,y)\,,\\
I_{n\textrm{-chain}}(Z)\equiv&\;\int_{z_1,...,z_n}\hat G(x,z_1)\times...\times \hat G(z_n,y)\,,\quad n\geq1\,.
\label{eq:Ichain}
\end{flalign}
This class of diagrams is particularly simple, as discussed in more detail below.

\begin{figure}[t]
\centering
\begin{subfigure}{0.5\textwidth}
\centering
\includegraphics[width=0.78\textwidth]{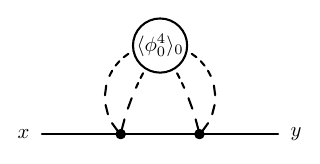}
\caption{}
\end{subfigure}%
\begin{subfigure}{0.5\textwidth}
\centering
\includegraphics[width=0.78\textwidth]{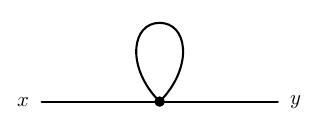}
\caption{}
\end{subfigure}
\caption{Diagrams at $\Lo(\lambda)$: the 2-chain diagram (a) and the non-zero mode tadpole diagram (b).}
\label{fig::Ediags2}
\end{figure}

Proceeding to $\Lo(\lambda)$ we have two diagrams: the 2-chain, shown in \figref{fig::Ediags2} (a), which corresponds to 
\begin{equation}
\bigg(-\frac{\lambda}{2}\bigg)^2\cor{}{\phi^4_0}_0\,I_{2\textrm{-chain}}(Z)\,,
\end{equation}
and the tadpole diagram involving the insertion of the $\lambda\hat\phi^4(x)$ vertex shown in \figref{fig::Ediags2} (b), corresponding to 
\begin{equation}
-\frac{\lambda}{2}I_{1\textrm{-chain}}(Z)\times\hat G(z,z)\,.
\end{equation}
We see from this expression that the coincident non-zero mode propagator $\hat G(z,z)$, which is a UV-divergent constant, factorises from the rest of the expression, which can be written in terms of the same integral we already encountered at $\Lo(\sqrt{\lambda})$, but with an $\Lo(\lambda)$-coefficient instead.  This diagram may therefore be regarded as a subleading correction to the $\Lo(\sqrt{\lambda})$-diagram, but since it lacks 
a $\ln |Z|$ enhancement, it contributes only to subleading logarithms, 
which we are not interested in here. The same argument holds for tadpole subgraphs in arbitrary diagrams. Therefore, we conclude that in the leading-logarithmic approximation diagrams with tadpole subgraphs 
never need to be considered.

\begin{figure}[t]
\centering
\begin{subfigure}{0.5\textwidth}
\centering
\includegraphics[width=0.8\textwidth]{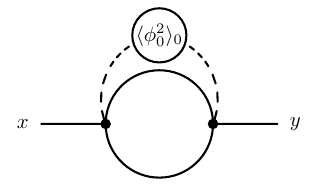}
\caption{}
\end{subfigure}%
\begin{subfigure}{0.5\textwidth}
\centering
\raisebox{0.7cm}[0pt][0pt]{\includegraphics[width=0.85\textwidth]{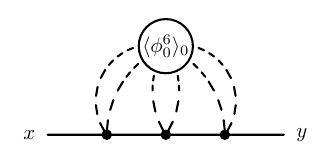}}
\caption{}
\end{subfigure}
\caption{The relevant $\Lo(\lambda^{\frac{3}{2}})$ diagrams: the non-zero-mode bubble diagram (a) and the 3-chain diagram (b).}
\label{fig::Ediags3}
\end{figure}

At $\Lo(\lambda^{\frac{3}{2}})$  a non-trivial non-zero-mode loop diagram appears for the first time. We can insert the $\lambda\phi_0\hat\phi^3(x)$ vertex twice to obtain the bubble diagram,  which is shown in \figref{fig::Ediags3} (a), and leads to
\begin{equation}
\frac{\lambda^2}{2}\cor{}{\phi^2_0}_0\,I_B(Z)\equiv\frac{\lambda^2}{2}\cor{}{\phi^2_0}_0\int_{z_1,z_2}\hat G(x,z_1)\hat G^2(z_1,z_2)\hat G(z_2,y)\,.
\label{eq::Ebubble}
\end{equation}
At the same order we also have the 3-chain, which is shown in \figref{fig::Ediags3} (b) and gives
\begin{equation}
\bigg(-\frac{\lambda}{2}\bigg)^3\cor{}{\phi^6_0}_0\,I_{3\textrm{-chain}}(Z)\,.
\end{equation}

\begin{figure}[t]
\centering
\begin{subfigure}{0.5\textwidth}
\centering
\includegraphics[width=0.8\textwidth]{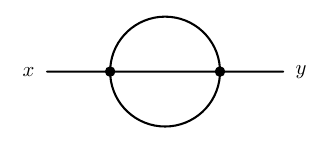}
\caption{}
\end{subfigure}%
\begin{subfigure}{0.5\textwidth}
\centering
\raisebox{0.1cm}[0pt][0pt]{\includegraphics[width=0.95\textwidth]{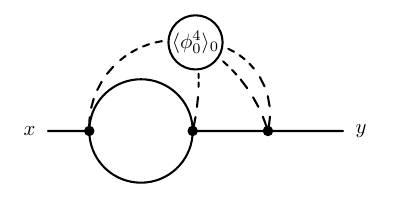}}
\caption{}
\end{subfigure}\\
\begin{subfigure}{0.5\textwidth}
\centering
\includegraphics[width=0.8\textwidth]{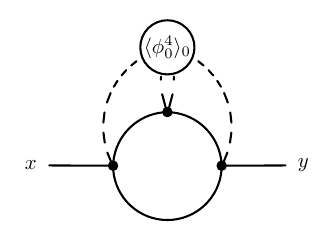}
\caption{}
\end{subfigure}%
\begin{subfigure}{0.5\textwidth}
\centering
\raisebox{0.75cm}[0pt][0pt]{\includegraphics[width=0.8\textwidth]{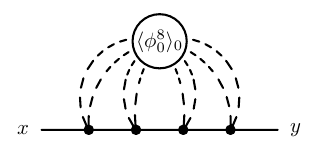}}
\caption{}
\end{subfigure}
\caption{The relevant $\Lo(\lambda^2)$ diagrams: the sunset diagram (a), the bubble-chain diagram (b), the bubble-chain-insertion diagram (c) and the 4-chain diagram (d).}
\label{fig:Ediags4}
\end{figure}

Finally, at $\Lo(\lambda^2)$ we have the four diagrams shown in \figref{fig:Ediags4}. The two-loop sunset diagram, shown in panel (a), leads to
\begin{equation}
\frac{\lambda^2}{6}I_S(Z)\equiv\frac{\lambda^2}{6}\int_{z_1,z_2}\hat G(x,z_1)\hat G^3(z_1,z_2)\hat G(z_2,y)\,,
\label{eq::Esunset}
\end{equation}
the bubble-chain diagram, shown in panel (b), plus its mirrored image along the vertical axis, leads to
\begin{equation}
-\frac{\lambda^3}{2}\cor{}{\phi^4_0}_0\,I_{BC}(Z)\equiv-\frac{\lambda^3}{2}\cor{}{\phi^4_0}_0\int_{z_1,z_2,z_3}\hat G(x,z_1)\hat G^2(z_1,z_2)\hat G(z_2,z_3)\hat G(z_3,y)\,,
\label{eq::EBC}
\end{equation}
the bubble diagram with a 1-chain insertion, shown in panel (c), to
\begin{equation}
-\frac{\lambda^3}{2}\cor{}{\phi^4_0}_0\,I_{BCI}(Z)=-\frac{\lambda^3}{2}\cor{}{\phi^4_0}_0\int_{z_1,z_2,z_3}\hat G(x,z_1)\hat G(z_1,z_2)\hat G(z_1,z_3)\hat G(z_3,z_2)\hat G(z_2,y)\,,
\label{eq::EBCI}
\end{equation}
and, the 4-chain, shown in panel (d), yields
\begin{equation}
\bigg(-\frac{\lambda}{2}\bigg)^4\cor{}{\phi^8_0}_0\,I_{4\textrm{-chain}}(Z)\,.
\end{equation}

\subsection{Extraction of the leading-logarithmic terms}

Having identified the diagrams of interest, we next extract the leading logarithm of each diagram in the limit of large invariant distance between the points $x$ and $y$ after analytic continuation to Lorentzian de Sitter space. The direct computation of Euclidean loop diagrams is very challenging \cite{Marolf:2010zp} and quickly becomes unfeasible, but the leading logarithm can be obtained relatively easily.

For the chain integrals the leading-logarithmic term can be determined in closed form. Details are presented in \appref{app::chainz}. The result reads
\begin{equation}
\lim\limits_{|Z|\rightarrow\infty}I_{n\textrm{-chain}}(Z)=-\frac{1}{(n+1)!}\frac{\log^{n+1}|Z|}{(d-1)^{n+1}H^{2n+2}V_d}\,, \quad n\in\N_0\,.
\label{eq::asychain}
\end{equation}
This formula is valid for both large time-like and space-like values of $Z$. For the diagrams involving non-zero-mode loops there is no general formula available, and we determine the leading-logarithmic terms on a case-by-case basis. 

For this purpose, we exploit the fact that the free non-zero mode propagator satisfies
\begin{equation}
\Box_x\hat G(x,y)\equiv\frac{1}{\sqrt{g(x)}}\,\p^x_{\mu}\Big[\sqrt{g(x)}g^{\mu\nu}(x)\p^x_{\nu}\hat G(x,y)\Big]=\frac{1}{V_d}-\frac{1}{\sqrt{g(x)}}\delta^{(d)}(x-y)\,,
\label{eq::Ghatdiffeq}
\end{equation}
where we used \eqref{eq::Gspec}, \eqref{eq::Yeigen} and \eqref{eq::Y0}. The same result holds if we replace $\Box_x\rightarrow\Box_y$ in the above equation. Acting with powers of $\Box_{x,y}$ on the diagrams  identified above replaces the external propagators with the right-hand side of \eqref{eq::Ghatdiffeq}. The delta-function term then removes  integrals associated to interaction vertices, effectively localising them at the external points. This is not so for the term proportional to $1/V_d$, but, as we will see below, these terms are always subleading. Further, the Laplace-Beltrami operator $\Box_x$ may be rewritten as~\cite{Spradlin:2001pw}
\begin{equation}
\Box_xf(Z)=H^2\bigg[(1-Z^2)\frac{\der^2}{\der Z^2}-d\cdot Z\frac{\der}{\der Z}\bigg]f(Z)
\end{equation}
when acting on a function $f$ which only depends on $Z$, which holds for all the above diagrams. The analytic continuation of this differential equation, and its solution, consists in trivially extending the range of values of $Z$ from the compact interval $[-1,1]$ to $(-\infty,\infty)$. The differential operator is invariant under $Z\rightarrow-Z$, and for $|Z|\rightarrow\infty$ it simplifies to
\begin{equation}
\lim\limits_{|Z|\rightarrow\infty}\Box_xf(Z)=-H^2\bigg[Z^2\frac{\der^2}{\der Z^2}+d\cdot Z\frac{\der}{\der Z}\bigg]f(Z)\,.
\end{equation}
Acting with powers of $\Box_{x,y}$ on the diagrams we identified above we obtain an inhomogeneous differential equation in $Z$ for them, where the inhomogeneous term can be related to simpler diagrams, and is therefore known. Taking the large-$|Z|$ limit of the analytically continued equation and solving it, allows us to extract the leading logarithm, sidestepping the full computation of the diagrams.

To illustrate and check this method we apply it to the 1-chain integral
\begin{equation}
I_{1\textrm{-chain}}(Z)=\int_z\;\hat G(x,z)\hat G(z,y)\,.
\end{equation}
By iterating the following procedure, the $n$-chain result can be recovered. 
Using \eqref{eq::Ghatdiffeq} and
\begin{equation}
\int_x\;\hat G(x,y)=0\,,
\label{eq::intGvanish}
\end{equation}
which follows from \eqref{eq::Yvanish}, we find
\begin{equation}
\Box_xI_{1\textrm{-chain}}(Z)=H^2\bigg[(1-Z^2)\frac{\der^2}{\der Z^2}-d\cdot Z\frac{\der}{\der Z}\bigg]I_{1\textrm{-chain}}(Z)=-\hat G(x,y)\,.
\end{equation}
In the limit $|Z|\rightarrow\infty$ this equation becomes
\begin{equation}
\bigg[-Z^2\frac{\der^2}{\der Z^2}-d\cdot Z\frac{\der}{\der Z}\bigg]I_{1\textrm{-chain},\infty}(Z)=\frac{\log|Z|}{(d-1)H^4V_d}\,,
\label{eq::asy1chain}
\end{equation}
where we used the known asymptotic behaviour of $\hat G(x,y)$,
\begin{equation}
\lim\limits_{|Z|\rightarrow\infty}\hat G(x,y)=-\frac{\log|Z|}{(d-1)H^2V_d}+\Lo(|Z|^0)\,,
\label{eq::gasy}
\end{equation}
and neglected subleading terms. The leading piece of the solution of \eqref{eq::asy1chain} reads
\begin{equation}
I_{1\textrm{-chain},\infty}(Z)=-\frac{\log^2|Z|}{2(d-1)^2 H^4V_d}+\Lo(\log|Z|)\,,
\label{eq::largeZ1chain}
\end{equation}
where here and in the following the subscript ``$\infty$" highlights that only the leading logarithmic term in the $|Z|\rightarrow\infty$-limit is kept. Eq.~\eqref{eq::largeZ1chain} reproduces the known leading asymptotic behaviour \eqref{eq::asychain} of the 1-chain diagram. 
In particular, since the differential equation obtained in this manner can be written as an inhomogeneous linear differential equation, where the inhomogeneity contains powers of $\log|Z|$, the leading logarithm of the result is unambiguous, since the integration constants affect only the subleading terms.

We next apply this method to the non-zero-mode loop diagrams identified in~\secref{sec::Ediaglist}.

\subsubsection{Bubble diagram}

We start with the bubble integral $I_B(Z)$, see \eqref{eq::Ebubble}. We let the operator $\Box_y\Box_x$ act on it and use \eqref{eq::Ghatdiffeq} to find
\begin{equation}
\Box_y\Box_xI_B(Z)=\hat G(x,y)^2-\frac{1}{V_d}\int_z\hat G(x,z)^2\,.
\label{eq::difeq}
\end{equation}
The last term in \eqref{eq::difeq} is constant, since by rotational invariance it cannot depend explicitly on $x$, and divergent. It contains the expected UV divergences generated by the loop integration, which can be removed as usual by renormalising the theory. The procedure leaves a renormalisation-scheme dependent finite constant in \eqref{eq::difeq}, which is subleading in the large-$|Z|$ limit compared to $\hat G(x,y)^2$. Therefore, we may ignore this term for the present discussion. This pattern will repeat itself in all the following computations, implying that the UV divergences and counterterm diagrams are irrelevant at large $|Z|$ in the leading-logarithmic approximation. This can be interpreted as the emergence of a semiclassical behaviour of the analytically continued Euclidean results in this limit, making them compatible with the stochastic results.

Thus keeping only the first term on the right-hand side of \eqref{eq::difeq}, taking the limit $|Z|\rightarrow\infty$, and using the large-$|Z|$ expression \eqref{eq::gasy} for $\hat G(x,y)^2$, we find the simpler differential equation
\begin{equation}
\bigg[-Z^2\frac{\der^2}{\der Z^2}-d\cdot Z\frac{\der}{\der Z}\bigg]^2I_{B\infty}(Z)=\frac{\log^2|Z|}{(d-1)^2H^8V^2_d}\,.
\end{equation}
Its leading-logarithmic solution reads 
\begin{equation}
I_{B\infty}(Z)=\frac{\log^4|Z|}{12(d-1)^4H^8V^2_d}+\Lo(\log^3|Z|)\,,
\label{eq::asybub}
\end{equation}
so for $|Z|\rightarrow\infty$ the leading term of the bubble diagram reads
\begin{equation}
\frac{\lambda^2}{2}\cor{}{\phi^2_0}_0\,I_{B\infty}(Z)=\frac{\lambda^2\cor{}{\phi^2_0}_0\log^4|Z|}{4!\,(d-1)^4H^8V^2_d}\,.
\end{equation}

\subsubsection{Sunset diagram}
We apply the same procedure to the sunset integral \eqref{eq::Esunset}, and find
\begin{equation}
\Box_y\Box_xI_S(Z)=\hat G(x,y)^3-\frac{1}{V_d}\int_z\hat G(x,z)^3\,.
\end{equation}
The last term on the right-hand side is divergent and constant, which makes it irrelevant for the leading large-$|Z|$ behaviour of the diagram. Thus, we only keep $\hat G(x,y)^3$ on the right-hand side in the above equation. Taking the large-$|Z|$ limit and using \eqref{eq::gasy} for $\hat G(x,y)^3$ we obtain
\begin{equation}
\bigg[-Z^2\frac{\der^2}{\der Z^2}-d\cdot Z\frac{\der}{\der Z}\bigg]^2I_{S\infty}(Z)=-\frac{\log^3|Z|}{(d-1)^3H^{10}V^3_d}\,.
\end{equation}
The leading solution reads
\begin{equation}
I_{S\infty}(Z)=-\frac{\log^5|Z|}{20(d-1)^5H^{10}V^3_d}+\Lo(\log^4|Z|)\,,
\end{equation}
hence the leading-logarithmic  term of the analytically continued sunset diagram is
\begin{equation}
\frac{\lambda^2}{6}I_{S\infty}(Z)=-\frac{\lambda^2\log^5|Z|}{5!\,(d-1)^5H^{10}V^3_d}\,.
\end{equation}

\subsubsection{Bubble-chain diagram}
Letting $\Box_y$ act on the bubble-chain integral, see \eqref{eq::EBC}, yields
\begin{equation}
\Box_yI_{BC}(Z)=-I_B(Z)\,,
\end{equation}
where \eqref{eq::intGvanish} was used. The leading asymptotic behaviour of $I_B(Z)$ is known from \eqref{eq::asybub}. Substituting it into the above equation we find the differential equation
\begin{equation}
\bigg[-Z^2\frac{\der^2}{\der Z^2}-d\cdot Z\frac{\der}{\der Z}\bigg]I_{BC\infty}(Z)=-\frac{\log^4|Z|}{12(d-1)^4H^{10}V^2_d}\,.
\end{equation} 
It has the leading solution
\begin{equation}
I_{BC\infty}(Z)=\frac{\log^5|Z|}{60(d-1)^5H^{10}V^2_d}+\Lo(\log^4|Z|)
\end{equation}
and the leading contribution in the $|Z|\rightarrow\infty$ limit of the bubble-chain diagram is then 
\begin{equation}
-\frac{\lambda^3}{2}\cor{}{\phi^4_0}_0I_{BC\infty}(Z)=-\frac{\lambda^3\cor{}{\phi^4_0}_0\log^5|Z|}{5!\,(d-1)^5H^{10}V^2_d}\,.
\end{equation}

\subsubsection{Bubble-chain-insertion diagram}
We let the operator $\Box_y\Box_x$ act on $I_{BCI}(Z)$, see \eqref{eq::EBCI}, and find
\begin{flalign}
\Box_y\Box_xI_{BCI}(Z)=&\;\hat G(x,y)\times I_{1\textrm{-chain}}(Z)-\frac{1}{V_d}\int_{z_1z_2}\hat G(x,z_1)\hat G(x,z_2)\hat G(z_1,z_2)\,.
\end{flalign}
The only relevant term in the leading large-$|Z|$ limit is the first one on the right-hand side, while the second one is, once again, constant and divergent. Taking $|Z|\rightarrow\infty$ and using \eqref{eq::gasy} and \eqref{eq::largeZ1chain} we find the differential equation
\begin{equation}
\bigg[-Z^2\frac{\der^2}{\der Z^2}-d\cdot Z\frac{\der}{\der Z}\bigg]^2I_{BCI\infty}(Z)=\frac{\log^3|Z|}{2(d-1)^3H^{10}V^2_d}\,.
\end{equation}
The leading-logarithmic solution reads
\begin{equation}
I_{BCI\infty}(Z)=\frac{\log^5|Z|}{40H^{10}(d-1)^5V^2_d}+\Lo(\log^4|Z|)
\end{equation}
and the contribution of the bubble-chain-insertion diagram, including the factor of 2 for the symmetric diagram, therefore reads
\begin{equation}
-\frac{\lambda^3}{2}\cor{}{\phi^4_0}_0I_{BCI\infty}(Z)=-\frac{3}{2}\frac{\lambda^3\cor{}{\phi^4_0}_0\log^5|Z|}{5!(d-1)^5H^{10}V^2_d}\,.
\end{equation}

\subsection{Summary and discussion}

\begin{table}[t]
\centering
\begin{tabular}{|c|l|}
\hline
\rule[-2ex]{0pt}{5ex}$\Lo(\lambda^{-\frac{1}{2}})$ &\quad $\cor{}{\phi^2_0}_0$\\
\hline
\rule[-3ex]{0pt}{7ex}$\Lo(\lambda^0)$ &\quad $\displaystyle -\frac{\log|Z|}{(d-1)H^2V_d}$\\
\hline
\rule[-3ex]{0pt}{7.5ex}$\Lo(\lambda^{\frac{1}{2}})$ &\quad $\displaystyle\frac{\lambda}{2}\cor{}{\phi^2_0}_0\frac{\log^2|Z|}{2!(d-1)^2H^4V_d}$\\
\hline
\rule[-3ex]{0pt}{7.5ex}$\Lo(\lambda)$ &\quad $\displaystyle-\frac{\lambda^2}{4}\cor{}{\phi^4_0}_0\frac{\log^3|Z|}{3!(d-1)^3H^6V_d}$\\
\hline
\rule[-3ex]{0pt}{7.5ex}$\Lo(\lambda^{\frac{3}{2}})$ &\quad $\displaystyle\bigg[\frac{\lambda^3}{8}\cor{}{\phi^6_0}_0+\frac{\lambda^2}{V_d}\cor{}{\phi^2_0}_0\bigg]\frac{\log^4|Z|}{4!(d-1)^4H^8V_d}$\\
\hline
\rule[-3ex]{0pt}{7.5ex}$\Lo(\lambda^2)$ &\quad $\displaystyle-\bigg[\frac{\lambda^4}{16}\cor{}{\phi^8_0}_0+\frac{5\lambda^3}{2V_d}\cor{}{\phi^4_0}_0+\frac{\lambda^2}{V^2_d}\bigg]\frac{\log^5|Z|}{5!(d-1)^5H^{10}V_d}$\\
\hline
\end{tabular}
\caption{Leading terms in large-$|Z|$ limit after analytic continuation.}
\label{tab::EuclargeZ}
\end{table}

The above results for the first six terms of the leading-log series \eqref{eq::GLLseries}  are summarised in \tabref{tab::EuclargeZ}. It is worth collecting two observations, which are valid to any order in $\sqrt{\lambda}$:
\begin{itemize}
\item Diagrams involving non-zero-mode tadpole subgraphs, as well as disconnected diagrams, contribute only at subleading order in the large-$|Z|$ limit after analytic continuation.
\item All effects from UV divergences and their renormalisation also contribute only subleading logarithms. 
\end{itemize}
In light of the complexity of the Euclidean diagrams, the simplicity of their leading-logarithmic behaviour is striking. It is also noteworthy that the dependence of a diagram on $\log|Z|$, $V_d$, and the combination $(d-1) H^2$, as well as its sign, appears to follow two simple rules: 
\begin{itemize}
\item Each non-coincident propagator $\hat G(x,y)$ in the diagram contributes a factor
\begin{equation}
-\frac{\log|Z|}{(d-1)H^2V_d}\,,
\end{equation}
which is the leading large-$|Z|$ limit of the propagator itself, see \eqref{eq::gasy}.
\item Each interaction vertex contributes a factor $V_d$, which can be interpreted as a remnant of the Euclidean spatial integration associated to each vertex.
\end{itemize}
The leading large-$|Z|$ behaviour can therefore be read off from any of the diagrams considered above by counting the number of non-coincident propagators. 

The above results can be summarised as follows:  diagrams containing 
tadpole subgraphs (which may themselves consist of only non-coincident propagators) do not contribute leading logarithms. At $\Lo(\lambda^{\frac{n}{2}})$ in the $\sqrt{\lambda}$-expansion of the Euclidean two-point function the remaining diagrams involve $n+1$ non-coincident propagators and contribute  leading-logarithmic terms of the form
\begin{equation}
(\sqrt{\lambda})^n\log^{n+1}|Z|
\label{eq::leadinglog}
\end{equation}
after analytic continuation. We explicitly checked the first six orders $0\leq n\leq 5$, and obtained a closed formula confirming this behaviour for arbitrary $n$ for the special case of the chain diagrams, see \eqref{eq::asychain}. This statement implies that the resummation of the chain diagrams proposed in \cite{Hollands:2011we} does not substantially ameliorate the behaviour of the analytically continued two-point function, since the neglected other diagram classes are equally  important. It seems plausible that the above two simple rules will continue to hold true to higher orders in the leading-logarithm series, and hence the conclusion before \eqref{eq::leadinglog}, but barring a proof we must leave this statement a conjecture. In particular, this implies that infinitely many classes of diagrams beyond the chain diagrams contribute to the leading-logarithmic series, which is compatible with the observation  \cite{LopezNacir:2018xto} that in the $O(N)$-symmetric version of the model infinitely many diagram classes contribute beyond the strict $N\rightarrow\infty$ limit. 

If the formulated conjecture is true, understanding the large-$|Z|$ behaviour of the analytically continued two-point function involves extracting the leading term in this limit from infinitely many Euclidean diagram classes, to arbitrary orders in the $\sqrt{\lambda}$ expansion, 
which does not appear feasible with the above method. 
However, we can still make progress on the large-$|Z|$ behaviour of the Euclidean two-point function by studying its stochastic counterpart.


\section{Asymptotic expansion of the stochastic two-point function}

We therefore switch to the stochastic formalism, and analyze the properties of the Taylor expansion of the stochastic two-point function \eqref{eq::gexp}, which is defined directly in Lorentzian de Sitter spacetime. 

\subsection{Agreement with fixed-order, analytically-continued Euclidean results}

The Taylor expansion of the autocorrelation function \eqref{eq::gexp} is the main point of contact between the stochastic and Euclidean formalisms.
Since the absolute value of $Z$ is large, we may approximate the argument of $\sG$ by
\begin{equation}
H^{-1}\log|2Z-1|\approx H^{-1}\log|Z|\,.
\label{eq::LLapprox}
\end{equation}
This approximation is valid at leading-logarithmic order, but the formalism based on the leading-order FP equation \eqref{eq::FP} is known to reproduce the full underlying field-theoretic description at this order (and only at this order), see e.g. \cite{Prokopec:2007akf}. Therefore, for our purposes no (relevant) information is lost through the replacement \eqref{eq::LLapprox}. Using \eqref{eq::taylorcoeff}, as well as 
\begin{equation}
\cor{}{\phi^{2n}_0}_0=\int_{-\infty}^{\infty}\der\vp\;\rho_{1\textrm{eq}}(\vp)\vp^{2n}\,,
\end{equation}
it is then immediate to verify that the coefficients of the stochastic series and the coefficients of the factors of $\log^n|Z|$ listed in \tabref{tab::EuclargeZ} agree up to, and including, $\Lo(\lambda^2)$:
\begin{equation}
\frac{1}{H^n}\frac{\der^n\sG(z)}{\der z^n}\bigg|_{z=0}=\frac{\der^nG(x,y)}{\der(\log|Z|)^n}\bigg|_{\log|Z|=0}\,,\quad 0\leq n\leq 5\,.
\label{eq:Eucstochequality}
\end{equation}
It may also be checked that for $d=4$ we reproduce the results given in section IV B of \cite{Starobinsky:1994bd}  up to the highest order $n=4$ given in that reference, using the relation 
\begin{equation}
\frac{\der^n\sG(z)}{\der z^n}\bigg|_{z=0}=\frac{3H^{2+n}}{2\pi\sqrt{\lambda}}\bigg(\frac{\lambda}{144\pi^2}\bigg)^{\!\frac{n}{2}}\frac{\der^n}{\der\tau^n}\bigg|_{\tau=0}g(\tau)\,,
\end{equation}
with $g(\tau)$ as defined in \cite{Starobinsky:1994bd}.

The correspondence between the perturbative Euclidean and stochastic results for the respective two-point functions seems to have gone unnoticed in the literature so far. It is noteworthy, because the stochastic formalism has been shown to faithfully reproduce the superhorizon behaviour of Lorentzian field-theory correlators \cite{Garbrecht:2014dca,Gorbenko:2019rza,Mirbabayi:2020vyt,Cohen:2021fzf}. The equivalence of the Euclidean and stochastic formalisms would thus imply the equivalence of the Euclidean and Lorentzian theories, at least in the superhorizon limit. The combined results of the works \cite{LopezNacir:2016gzi,LopezNacir:2018xto,LopezNacir:2019ord} and \cite{Gorbenko:2019rza,Gautier:2013aoa,Gautier:2015pca} show this to be true for the two-point function in the $O(N)$ vector model at leading and next-to-leading order in the $1/N$ expansion. The results found above provide evidence that this may be true also for the single-field case.

\subsection{Large-order behaviour}

The natural question that follows from the found fixed-order, leading-logarithmic equivalence of the Euclidean and stochastic results is whether the series in question may be summed to yield the full answer. This is a non-trivial question from the Euclidean point of view, since perturbative expansions in QFT are typically asymptotic series, rather than convergent ones. Thus, it is interesting to analyze the large-order behaviour of the stochastic series, since it might allow us to form a hypothesis about the physical content of its Euclidean counterpart. A related question was studied in \cite{Honda:2023unh}, where the large-order behaviour of the time-dependent non-equilibrium stochastic one-point functions $\cor{}{\bar\phi^n(t)}$ was examined. 
It was found numerically that the Taylor expansions of the one-point functions around $t=0$ are divergent, and Pad\'e-Borel resummation provided numerical evidence that these divergent expansions can nevertheless be summed to recover the exact results. In the following we study the stochastic equilibrium two-point function, with a focus on analytic methods and results.

In \cite{Starobinsky:1994bd} it is remarked without proof that the series in question is asymptotic, with zero radius of convergence. 
We now show that this is indeed the case. To examine its convergence properties it is convenient to rewrite the Taylor expansion \eqref{eq::gexp} as 
\begin{equation}
\sG(z)\equiv\frac{H^{\frac{d}{2}}}{\sqrt{\lambda}}\sum_{n=0}^{\infty}\sG_n\Big(H^{\frac{d}{2}-1}\sqrt{\lambda}z\Big)^{\!n} \equiv \frac{H^{\frac{d}{2}}}{\sqrt{\lambda}}\sum_{n=0}^{\infty}\sG_n\tau^n\,.
\label{eq::Gexpdimless}
\end{equation}
This defines the dimensionless variable $\tau=H^{\frac{d}{2}-1}\sqrt{\lambda}z$, and the dimensionless Taylor coefficients
\begin{equation}
\sG_n\equiv\frac{1}{n!}\int_{-\infty}^{\infty}\der\tilde\vp\;\tilde\rho_{1\textrm{eq}}(\tilde\vp)\tilde\vp(\hat\Gamma_{\tilde\vp})^n\tilde\vp\,,
\label{eq::Gdimless}
\end{equation}
which are in turn computed in terms of the dimensionless quantities
\begin{equation}
\tilde\vp\equiv\bigg(\frac{\lambda}{H^d}\bigg)^{\!\frac{1}{4}}\vp\,,\quad \tilde\rho_{1\textrm{eq}}(\tilde\vp)\equiv\bigg(\frac{H^d}{\lambda}\bigg)^{\!\frac{1}{4}}\rho_{1\textrm{eq}}(\tilde\vp)\,,\quad \hat\Gamma_{\tilde\vp}\equiv\frac{1}{d-1}\bigg[-\frac{1}{6}\tilde\vp^3\frac{\p}{\p\tilde\vp}+\frac{1}{V_dH^d}\frac{\p^2}{\p\tilde\vp^2}\bigg]\,.
\label{eq::stochdimless}
\end{equation}
The $\sG_n$ are related to the Taylor coefficients in \eqref{eq::gexp} via
\begin{equation}
H^{\frac{d}{2}(n+1)-n}(\sqrt{\lambda})^{n-1}\sG_n=\frac{1}{n!}\frac{\der^n\sG(z)}{\der z^n}\bigg|_{z=0}\,.
\end{equation}
To check the convergence of the series we may use the ratio test: the series diverges if for arbitrary fixed $\tau\neq0$
\begin{equation}
L\equiv\lim\limits_{n\rightarrow\infty}\tau\,\bigg|\frac{\sG_{n+1}}{\sG_n}\bigg|>1\,.
\label{eq::rtest}
\end{equation}
The leading estimate for the large-order behaviour of the Taylor coefficients is derived in \appref{app::Gn}, and the result is
\begin{equation}
\lim\limits_{n\rightarrow\infty}\sG_n\approx A\,C^n\,\frac{(2n-1)!!}{n!}\Gamma\bigg(\frac{2n+3}{4}\bigg)\,,\quad A\,,C\in\R\,.
\label{eq:Gnasy}
\end{equation}
The constants $A$ and $C$, while assumed to be non-vanishing, are otherwise undetermined so far. Plugging this estimate into \eqref{eq::rtest} we find
\begin{equation}
L=|C|\tau\lim\limits_{n\rightarrow\infty}\Big[\sqrt{2n}+\Lo(n^{-\frac{1}{2}})\Big]\rightarrow\infty
\label{eq::rtestres}
\end{equation}
for any non-zero $C$ and $\tau$. The obtained result indeed grows like $\sqrt{n}$ for large $n$, showing that the radius of convergence of the series is zero, as stated in \cite{Starobinsky:1994bd}. The Taylor series of $\sG(z)$ around $z=0$ is thus a divergent power series of Gevrey-$\frac{1}{2}$ class. 

\begin{figure}[t]
\centering
\begin{subfigure}{0.5\textwidth}
\centering
\includegraphics[width=0.97\textwidth]{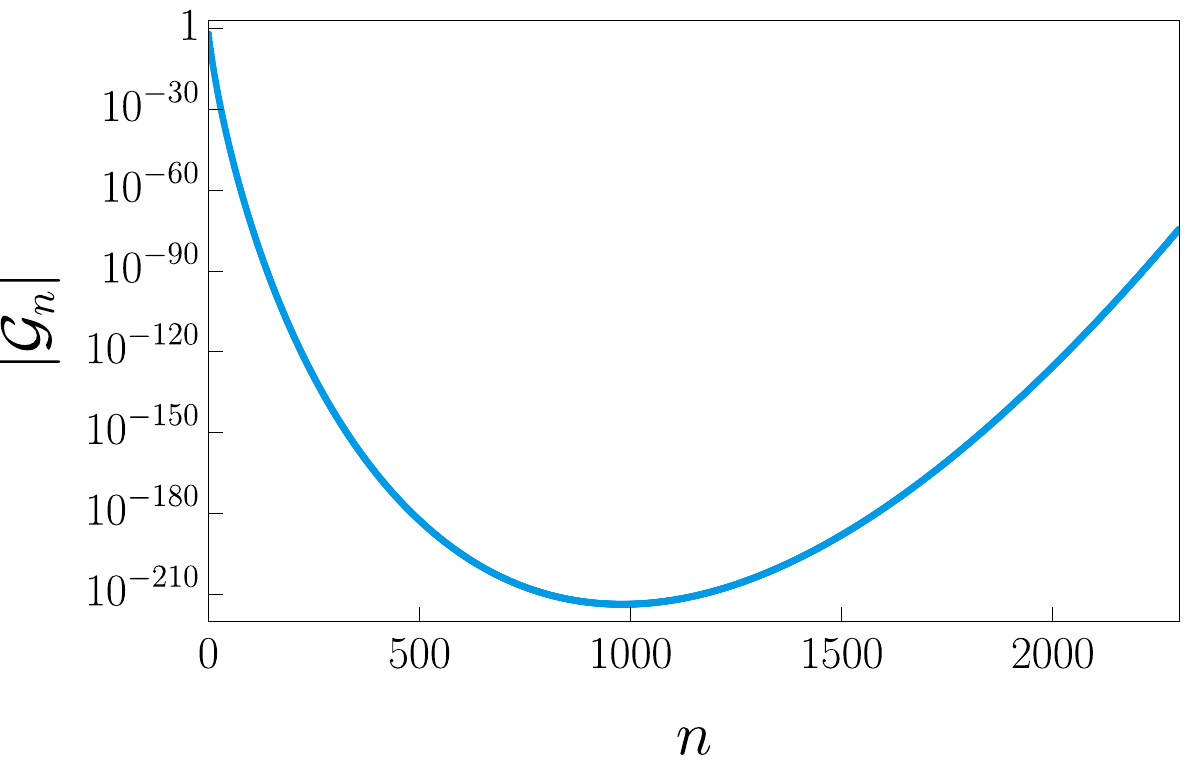}
\caption{}
\end{subfigure}%
\begin{subfigure}{0.5\textwidth}
\centering
\includegraphics[width=0.97\textwidth]{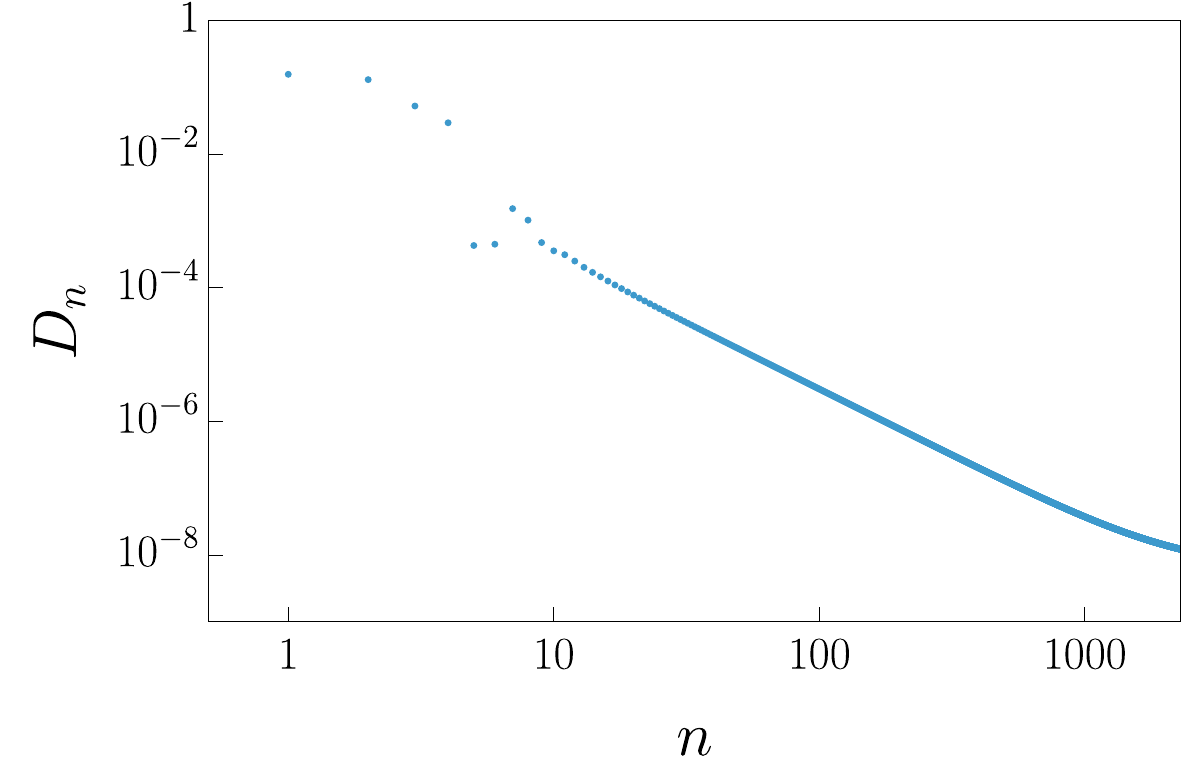}
\caption{}
\end{subfigure}
\caption{(a) Log-linear plot of the absolute value of the dimensionless Taylor coefficients $\sG_n$ plotted against $n\in[0,2300]$ and (b) log-log plot of the absolute value of the deviation of the ratio of adjacent Taylor coefficients and the formula $f(n)=C\sqrt{2n}$ from 1, defined as $D_n\equiv \big||(\sG_{n+1}/\sG_n)\big/(C\sqrt{2n})|-1\big|$, plotted against $n\in[1,2300]$.}
\label{fig::coeffplot}
\end{figure}

We next check this behaviour numerically for the physically relevant value $d=4$, and determine the constants $A$ and $C$. To this end, we compute 
the coefficients $\sG_n$ by evaluating \eqref{eq::Gdimless} for $n\in[0,2300]$ using {\asciifamily Mathematica}. Their absolute values as a function of $n$ are plotted in \figref{fig::coeffplot} (a). They exhibit the typical pattern of an asymptotic series, with optimal truncation at $n\approx1000$. The $\sG_n$ are sign-alternating, starting from $\sG_0>0$, a pattern that can already be seen in \tabref{tab::EuclargeZ}, and that holds to all orders due to \eqref{eq::taylorcoeff2}, since $b_n>0$ and $\Lambda_n>0$. The ratio of adjacent coefficients may be fitted by the function 
\begin{equation}
f(n)=(0.031951573...)\,\sqrt{n}\,.
\label{eq::fitfct}
\end{equation}
As may be seen in \figref{fig::coeffplot} (b) 
the absolute value of the ratio $|\sG_{n+1}/\sG_n|$ converges very quickly to the expected square-root form, the error between the points and the fit being of $\Lo(10^{-7})$ already at $n\approx200$, and decreasing after that point. Comparing \eqref{eq::rtestres} at $\tau=1$ to \eqref{eq::fitfct} we find the numerical value
\begin{equation}
C=-0.022593174...\,.
\label{eq:Cnum}
\end{equation}
The overall constant $A$ can be determined by taking the ratio of the estimated $\sG_n$ with the numerically computed true values of $\sG_n$ for large values of $n$. Requiring it to equal 1 yields 
\begin{equation}
A=0.42102...\,.
\label{eq:Anum}
\end{equation}
We thus find the leading estimate for the Taylor coefficients
\begin{equation}
\lim\limits_{n\rightarrow\infty}\sG_n\approx (0.42102...)\times (-0.022593174...)^n
\,\frac{(2n-1)!!}{n!}\,\Gamma\bigg(\frac{2n+3}{4}\bigg)\,.
\label{eq::Gestimate}
\end{equation}

As a further check on this result, we use it to extract the location and magnitude of the minimal series coefficient $\sG_{n_{\textrm{min}}}$, and compare the result with \figref{fig::coeffplot} (a). Applying the Stirling approximation we find the alternative expression 
\begin{equation}
\lim\limits_{n\rightarrow\infty}\sG_n=2^{\frac{1}{4}}A\,e^{-\frac{n}{2}}\,(2C^2n)^{\frac{n}{2}}\Big[1+\Lo(n^{-1})\Big]
\label{eq::estimatestandardform}
\end{equation}
 for $\sG_n$ for large $n$. The above expression has a minimum at
\begin{equation}
n_{\textrm{min}}=\frac{1}{2C^2}\approx 980\,,
\end{equation}
with value
\begin{equation}
\sG_{n_{\textrm{min}}}\approx 10^{-213}\,,
\end{equation}
in good agreement with \figref{fig::coeffplot} (a).

An analytic expression for the constants $A$ and $C$ may be obtained from the WKB approximation\footnote{The WKB analysis was carried out with the help of Claude Opus 5.0 and 5.5.} for the asymptotic behaviour of the eigenvalues and eigenfunctions $\Lambda_n$, $\Phi_n$ defined by \eqref{eq::Schrodi}, together with the formulae \eqref{eq::bn} and \eqref{eq::taylorcoeff2}. They read 
\begin{flalign}
A&=\frac{12\cdot 3^{\frac{3}{8}}(2\pi)^{\frac{3}{4}}}{\Gamma(\frac{1}{4})\Gamma(\frac{1}{3})^{\frac{9}{4}}\sqrt{V_dH^d}}\,,\label{eq:A}\\
C&=-\frac{4\cdot3^{\frac{1}{4}}\pi^{\frac{3}{2}}}{\Gamma(\frac{1}{3})^{\frac{9}{2}}(d-1)\sqrt{V_dH^d}}\,.\label{eq:C}
\end{flalign}
The derivation is presented in \appref{app:ACderivation}. For $d=4$ these formulae reproduce the numerically determined values for $A$ and $C$ shown in \eqref{eq:Cnum} and \eqref{eq:Anum}.\footnote{By adding $1/n$ and $1/n^2$ terms to \eqref{eq:Gnasy} and fitting their coefficients to the first 2400 terms, we obtained numerical expressions 
for $A$ and $C$ with several more digits than given in \eqref{eq:Cnum}, \eqref{eq:Anum}, which all agree with the presented analytic forms.} 

\subsection{Borel summability}

The next natural question is whether the exact solution for the two-point function \eqref{eq:curlyG} can be reconstructed from its asymptotic expansion by  Borel summation. The  Borel transform of the series  expansion of $\sG(z)$ is defined as 
\begin{equation}
B[\sG](u)\equiv\sum_{n=0}^{\infty}\frac{\sG_n}{n!}\,u^n\,,
\label{eq::borelt}
\end{equation} 
and the Borel sum $\sG_B(z)$ of the asymptotic series (accounting for the normalisation of \eqref{eq::Gexpdimless}) is then given by
\begin{equation}
\sG_B(z)\equiv \frac{H^{\frac{d}{2}}}{\sqrt{\lambda}}\frac{1}{\tau}\int_0^{\infty}\der u\;e^{-\frac{u}{\tau}}B[\sG](u)\,.
\label{eq::borels}
\end{equation}
If the series is Borel-summable, the integral exists and $\sG_B(z)$ coincides with  the exact result~$\sG(z)$. 

To prove Borel summability of the series, we check the assumptions of the Nevanlinna-Sokal theorem \cite{Sokal:1980ey}, which is stated in \appref{app:NStheorem}. This requires the knowledge of the asymptotic behaviour of the series coefficients $\sG_n$, and of the domain of analyticity of the original function $\sG(z)$. 

From the large-$n$ behaviour of the Taylor coefficients \eqref{eq::estimatestandardform} and the found values for $A$ and $C$ \eqref{eq:Cnum}, \eqref{eq:Anum}, we can read off that the series of interest is a sign-alternating asymptotic series of Gevrey-$\frac{1}{2}$ type. We may thus estimate the remainder of the difference of $\sG(z)$ and the $(N-1)$-st partial sum of its Taylor expansion around $z=0$ as
\begin{equation}
\left|\,\sG(z)-\frac{H^{\frac{d}{2}}}{\sqrt{\lambda}}\sum_{n=0}^{N-1}\sG_n\tau^n\right|\leq a\,c^{N}\sqrt{N!}\,|z|^{N}
\label{eq:Gexpbound}
\end{equation}
for some $a,\,c>0$. The Nevanlinna-Sokal theorem requires only a much weaker $N!$ 
bound for the remainder, which is amply satisfied by the above inequality. To determine the domain of analyticity of $\sG(z)$ we inspect its Taylor expansion around an arbitrary point $z_0\in\C$:
\begin{equation}
\sG(z)=\sum_{n=0}^{\infty}\frac{1}{n!}\frac{\der^n\sG(z)}{\der z^n}\bigg|_{z=z_0}(z-z_0)^n\,.
\label{eq::TaylorGz0}
\end{equation}
Writing down such a series expansion for the case at hand explicitly involves the interchange of two infinite sums, as can be seen from \eqref{eq:curlyG} using the series representation of the exponential function:
\begin{equation}
\sG(z)=\sum_{n=0}^{\infty}b_n\sum_{m=0}^{\infty}\frac{1}{m!}(-\Lambda_n)^me^{-\Lambda_nz_0}\,(z-z_0)^m\stackrel{?}{=}\sum_{m=0}^{\infty}\frac{1}{m!}\left[\sum_{n=0}^{\infty}b_n(-\Lambda_n)^me^{-\Lambda_nz_0}\right](z-z_0)^m\,.
\end{equation}
By Fubini's theorem, this interchange preserves the value of the double sum only if it is absolutely convergent,
\begin{equation}
\sum_{n=0}^{\infty}\sum_{m=0}^{\infty}\bigg|\frac{b_n}{m!}(-\Lambda_n)^me^{-\Lambda_nz_0}(z-z_0)^m\bigg|=\sum_{n=0}^{\infty}b_ne^{-\Lambda_n[\Re(z_0)-|z-z_0|]}\,<\infty\,,
\end{equation}
where we have used that $b_n>0$, $\Lambda_n>0$. The sum on the right-hand side converges for
$\Re(z_0)-|z-z_0|\geq 0$, 
which can only be satisfied for $\Re(z_0)\geq0$. The boundary case $\Re(z_0)-|z-z_0|=0$ is included, because the sum over the $b_n$ converges, see \eqref{eq::bn}. Decomposing $z$ and $z_0$ into real and imaginary parts, $z_{(0)}=x_{(0)}+iy_{(0)}$, this inequality translates into 
\begin{equation}
\sqrt{(x-x_0)^2+(y-y_0)^2}\leq x_0\,,
\label{eq:analyticdom}
\end{equation}
which defines a closed disc with radius $x_0$ centered at the coordinates $(x_0,y_0)$ in the complex plane. Thus, for values of $z$ which lie within this region, the Taylor expansion \eqref{eq::TaylorGz0} is well-defined and converges to the same result as the original function, which means that the function is analytic in this disc. The point $z_0$ is arbitrary, so we may use the disc regions defined by \eqref{eq:analyticdom} to cover the entire complex half-plane with $\Re(z)>0$, and this defines the domain of analyticity of $\sG(z)$. Notice that the imaginary axis is excluded. Therefore, even though $\sG(z)$ is infinitely-often differentiable at $z=0$, its Taylor expansion at that point is not expected to converge. Given \eqref{eq:Gexpbound}, the Nevanlinna-Sokal theorem only requires $\sG(z)$ to be holomorphic in an open disc $C_R=\{z\,|\,\Re(z^{-1})>R^{-1}\}$ tangent to the imaginary axis for Borel summability, and to reproduce the original function from its divergent series representation. The domain of analyticity of $\sG(z)$ is therefore more than large enough for the theorem to hold. We may thus conclude that the asymptotic series expansion of $\sG(z)$ at the point $z=0$ is Borel-summable, the Borel sum $\sG_B(z)$ \eqref{eq::borels} converges absolutely for any $z$ with $\Re(z)>0$, and it reproduces the original function $\mathcal{G}(z)$. 

Indeed, we can show the Borel summability of the series \eqref{eq::Gexpdimless} directly. The Borel transform \eqref{eq::borelt} converges absolutely, and its radius of convergence is infinite,
\begin{equation}
\sum_{n=0}^{\infty}\frac{|\sG_n|}{n!}|u|^n<\infty\,,
\end{equation}
for any $u$. We can check this statement using the ratio test and the known leading asymptotic behaviour of the Taylor coefficients \eqref{eq::Gestimate}:
\begin{flalign}
\lim\limits_{n\to\infty}\frac{n!|\sG_{n+1}||u|^{n+1}}{(n+1)!|\sG_n||u|^n}=\lim\limits_{n\to\infty}|C||u|\,\sqrt{\frac{2}{n}}=0\,.
\end{flalign}
Therefore, using 
\begin{equation}
\sG_n=\frac{1}{n!}\sum_{k=0}^{\infty}\beta_k(-\lambda_k)^n\,,
\end{equation}
with the dimensionless quantities
\begin{equation}
\beta_n\equiv\frac{\sqrt{\lambda}}{H^{\frac{d}{2}}}\,b_n\,,\quad \lambda_n\equiv\frac{\Lambda_n}{H^{\frac{d}{2}-1}\sqrt{\lambda}}\,,
\end{equation}
we may write
\begin{equation}
B[\sG](u)=\sum_{n=0}^{\infty}u^n\sum_{k=0}^{\infty}\frac{1}{(n!)^2}\,\beta_k(-\lambda_k)^n=\sum_{k=0}^{\infty}\beta_k\sum_{n=0}^{\infty}\frac{1}{(n!)^2}(-\lambda_k)^nu^n=\sum_{k=0}^{\infty}\beta_k J_0\left(2\sqrt{\lambda_ku}\right)\,,
\label{eq:doublesuminterchange}
\end{equation}
where $J_{\nu}(z)$ denotes the Bessel function of the first kind, and the exchange of the two sums is justified by Fubini's theorem. The Borel sum \eqref{eq::borels} is then readily evaluated,
\begin{equation}
\sG_B(z)=\frac{H^{\frac{d}{2}}}{\sqrt{\lambda}}\sum_{k=0}^{\infty}\frac{\beta_k}{\tau}\int_0^{\infty}\der u\;e^{-\frac{u}{\tau}}J_0\Big(2\sqrt{\lambda_ku}\Big)=\sum_{k=0}^{\infty}b_ke^{-\Lambda_kz}=\sG(z)\,,
\end{equation}
where the exchange of the sum and integral is, once again, justified by the absolute convergence of $B[\sG](u)$ and of the above integral, which is guaranteed by the Nevanlinna-Sokal theorem. Thus, we have explicitly shown that Borel summation yields the original function from its divergent series representation at $z=0$.

Working within the stochastic formalism, the Borel summability of the expansion of the two-point function is essentially a 
mathematical curiosity, since one can always resort to solving the Fokker-Planck equation numerically to obtain
the exact solution. However, when viewed through the Euclidean lens, this observation becomes highly non-trivial and potentially useful, since in that setting the exact solution of the two-point function is not available. It raises the possibility that the analytically continued Euclidean two-point function may be summed in the same manner, which would imply that the large-$|Z|$ behaviour of the Lorentzian result is encoded in its Euclidean counterpart. The equality of the fixed-order results found in \secref{sec::Euc2pt} and their stochastic counterparts is encouraging. To rigorously prove this hypothesis, an all-order understanding of the leading large-$|Z|$ behaviour of the terms in the expansion of the Euclidean result would be needed. 

\subsection{Extracting the $\Lambda_n$ and $b_n$ from low-order Taylor coefficients}
\label{sec::coeffmatch}

While proving the Borel summability of the series expansion of $\sG$ we have assumed that all its expansion coefficients are known. However, in practice only a finite number of them can be computed analytically, since a closed-form expression for the coefficients $\sG_n$ for arbitrary $n$ is not available for the  massless $\lambda\phi^4$-theory. Furthermore, we have only proven the equivalence of the expansion of the Euclidean and stochastic two-point functions for $0\leq n\leq 5$, see \eqref{eq:Eucstochequality}. To better understand to which extent the  behaviour of the Lorentzian two-point function can be extracted from Euclidean data, we may therefore also ask whether the lowest relaxation eigenvalues $\Lambda_n$ and coefficients $b_n$ can be determined, given only the knowledge of the functional form of $\sG$ \eqref{eq:curlyG} and the first few expansion coefficients of the Euclidean two-point function. We show that this is possible using the formula \eqref{eq::taylorcoeff2} for the Taylor coefficients, which we reprint here for convenience:
\begin{equation}
\frac{\der^n\sG(z)}{\der z^n}\bigg|_{z=0}=\sum_{m=0}^{\infty}b_m(-\Lambda_m)^n\,.
\label{eq:coeff}
\end{equation}
This infinite sum converges for any $n$, 
and from the large-$m$ asymptotics of the $\Lambda_m$ and $b_m$, see \appref{app:ACderivation}, we know that the summands decrease exponentially in magnitude for large values of $m$. Thus, we may truncate the series in \eqref{eq:coeff} at some finite $m$ in a controlled fashion. Given the first $2n$ Taylor coefficients of $\sG$, the truncations of the corresponding $2n$ series at $m=n$ define a set of $2n$ equations for $2n$ variables, which may be uniquely solved for the lowest $n$ pairs of $\Lambda_n$ and $b_n$. 
Making use of the first six coefficients given in \tabref{tab::EuclargeZ}, this leads to the values listed in the upper table in \tabref{tab::Lb}. 

\begin{table}[t]
\centering
\begin{tabular}{|c|c|c|c|c|}
\hline
$n$ & \multicolumn{2}{c|}{$\Lambda_{n,\textrm{ approx}}$} & \multicolumn{2}{c|}{$b_{n,\textrm{ approx}}$}\\
\hline
1 & \multicolumn{2}{c|}{$0.0363302450H\sqrt{\lambda}$}& \multicolumn{2}{c|}{$0.3178731733H^2/\sqrt{\lambda}$}\\
3 & \multicolumn{2}{c|}{$0.2250452972H\sqrt{\lambda}$} & \multicolumn{2}{c|}{$0.0048323915H^2/\sqrt{\lambda}$}\\
5 & \multicolumn{2}{c|}{$0.5814723498H\sqrt{\lambda}$} &  \multicolumn{2}{c|}{$0.0000502700H^2/\sqrt{\lambda}$}\\
\hline
\hline
$n$ & $\Lambda_n$ & $\Delta\Lambda_n$ in \% & $b_n$ & $\Delta b_n$ in \%\\
\hline
1 & $0.0363030441H\sqrt{\lambda}$ & $0.08$ & $0.3177346504H^2/\sqrt{\lambda}$ & $\phantom{-}0.04$\\
3 & $0.2190951808H\sqrt{\lambda}$ & $2.72$ & $0.0049057183H^2/\sqrt{\lambda}$ & $-1.50$ \\
5 & $0.4736902674H\sqrt{\lambda}$ & $22.75$ & $0.0001126695H^2/\sqrt{\lambda}$ & $-55.38$\\
\hline 
\end{tabular}
\caption{Approximate values of the relaxation eigenvalues $\Lambda_n$ and coefficients $b_n$ obtained from the first six Taylor coefficients of $\sG$ (upper table), and numerically determined values of $\Lambda_n$ and $b_n$ together with their deviation $\Delta\Lambda_n\equiv(\Lambda_{n,\textrm{ approx}}-\Lambda_n)/\Lambda_n$ and $\Delta b_n\equiv (b_{n,\textrm{ approx}}-b_n)/b_n$ from the approximate values (lower table).}
\label{tab::Lb}
\end{table}

We may compare these results with the numerical values for the $\Lambda_n$ and $b_n$, given in the lower table in \tabref{tab::Lb}, which were determined by solving \eqref{eq::Schrodi} numerically. This is done by guessing a value for a given $\Lambda_n$, solving \eqref{eq::Schrodi} numerically for the corresponding eigenfunction $\Phi_n(\vp)$, and then adjusting the guessed value of $\Lambda_n$ to improve the convergence towards zero of the numerical solution for $\Phi_n(\vp)$ for large values of $\vp$, up to the desired accuracy. By counting the number of zero-crossings of the resulting $\Phi_n(\vp)$, and using the node theorem for the Schr\"odinger equation, one then determines to which $n$ the determined $\Lambda_n$ and $\Phi_n(\vp)$ correspond.  Using the resulting numerical solution for the corresponding eigenfunctions $\Phi_n$, the coefficients $b_n$ 
as defined in \eqref{eq::bn} are then computed by numerical integration. 
The values we obtain this way agree with the ones reported in \cite{Markkanen:2019kpv} (after adjusting for the different normalisation of the quartic coupling) for $n=1,3$, as the highest order considered in that reference is $n=4$, and up to the accuracy given in the reference. 

We see that the approximate values of the $b_n$ undershoot the numerically determined ones for $n=3,5$, while the approximate values of $\Lambda_n$ consistently overshoot them. The accuracy of the approximation for both $b_n$ and $\Lambda_n$ deteriorates rapidly as $n$ increases. Nevertheless, $b_1$ and $\Lambda_1$, which is the pair that determines the leading large-$|Z|$ behaviour of $\sG$, are obtained to sub-permille accuracy with only the first six Taylor coefficients, which we have explicitly shown to agree with the analytically continued Euclidean result. Therefore, even given the knowledge of the first few leading-logarithmic terms of the $\sqrt{\lambda}$-expansion of the Euclidean two-point function, the leading large-$|Z|$ behaviour of its Lorentzian counterpart can be determined quite accurately. This simple method may, in principle, be pushed beyond the leading-logarithmic approximation, to determine the subleading terms of the large-$|Z|$ behaviour of the Lorentzian two-point function. 


\section{Conclusion}

We studied the analytic continuation of the two-point function $\cor{}{\phi(x)\phi(y)}$ in massless $\lambda\phi^4$-theory in Euclidean de Sitter space, in the limit of large invariant distance $Z_{xy}$, and the asymptotic behaviour and Borel summability of the stochastic two-point function $\sG(z)$. 
The starting point for the Euclidean analysis is a non-perturbative treatment of the zero mode, which self-regularises the infrared divergence present in Euclidean momentum integrals through ``dynamical mass'' generation. The ensuing $\sqrt{\lambda}$-expansion still contains logarithms of the distance, which turn into untamed secular logarithms upon analytic continuation to Lorentzian de Sitter space. These are summed by the stochastic formalism, which has no analogue in Euclidean de Sitter space. 

We determined the leading terms in the large-distance limit of the two-point function in Euclidean space, which are powers of the logarithm $\log|Z_{xy}|$, up to N$^5$LO in the $\sqrt{\lambda}$-expansion. We then showed that the result for the Taylor expansion of the stochastic equilibrium two-point function around $z=0$ is reproduced up to this order. By determining its expansion around the coincidence limit to very high order numerically, as well as the leading asymptotic behaviour of the relaxation eigenvalues $\Lambda_n$ and two-point function coefficients~$b_n$,
\begin{flalign}
\lim\limits_{n\to\infty}\Lambda_n&=\frac{16\pi^3\sqrt{\lambda}}{9\Gamma(\frac{1}{3})^{\frac{9}{2}}H(d-1)\sqrt{V_d}}\,n^{\frac{3}{2}}\,,\\
\lim\limits_{n\to\infty}b_n&=2^{\frac{1}{4}}\frac{16\cdot 3^{\frac{3}{4}}\pi^{\frac{3}{2}}}{\Gamma(\frac{1}{4})\Gamma(\frac{1}{3})^{\frac{9}{4}}\sqrt{V_d\lambda}}\,n^{-\frac{1}{4}}\exp\left(-\frac{\pi}{\sqrt{3}}n\right),
\end{flalign}
we analysed the large-order behaviour of the Taylor expansion of $\sG(z)$, showed it to be a sign-alternating asymptotic series of Gevrey-$\frac{1}{2}$ type, and proved its Borel summability.  Finally, we demonstrated that its leading large-$z$ behaviour can be accurately determined from the knowledge of its functional form and the first few series coefficients.

Even though the Euclidean results are limited to fixed-order perturbation theory, they are suggestive enough to lead us to formulate the following two hypotheses:
\begin{enumerate}[label=\arabic*)]
\item Diagrams containing tadpole subgraphs do not contribute leading logarithms. At $\Lo(\lambda^{\frac{n}{2}})$ in the $\sqrt{\lambda}$-expansion of the Euclidean two-point function the remaining diagrams involve $n+1$ non-coincident propagators and contribute  leading-logarithms proportional to $(\sqrt{\lambda})^{n-1}\log^n|Z_{xy}|$ in the large-$|Z_{xy}|$ limit after analytic continuation.
\item The analytic continuation of the interacting Euclidean two-point function to the Lorentzian setting yields a well-defined expansion in powers of $\sqrt{\lambda}$ of the Lorentzian correlator computed in a non-perturbative vacuum state which respects the full set of de Sitter isometries. The leading-logarithmic
terms in the $|Z_{xy}|\rightarrow\infty$ limit of this result agree with the stochastic expansion of this quantity to all orders,
and can be summed using Borel summation to obtain a well-behaved two-point function for large space-like or time-like values of $Z_{xy}$.
\end{enumerate}
It is suggestive and intriguing from the results of the present work that the $\sqrt{\lambda}$-expansion of the Euclidean two-point function might be analytically continued and Borel-summed 
to the stochastic solution. If proven to be true, these hypotheses would settle the physical meaning of the Euclidean prescription to compute massless correlation functions in the $\sqrt{\lambda}$-expansion, at least in the case of the two-point function, while also providing a procedure to sum the large logarithmic terms resulting from the analytic continuation away from the sphere. 

Beyond settling these open points, it would be interesting to look for further connections between the Euclidean and stochastic results. The Euclidean setting is naturally suited to study questions about effectively strongly-coupled QFT in de Sitter space \cite{DiPietro:2023inn} posed at small invariant distances, rather than the superhorizon limit that is usually of interest for inflationary physics. Clarifying its status could make it a useful tool to explore such questions. One particularly interesting direction could be to understand the physical meaning of the relaxation eigenvalues, which remains an unaddressed topic in the literature. Finding a field-theoretic interpretation for them, for instance analysing the K\"all\'en-Lehmann representation \cite{Bros:1995js,Loparco:2023rug} of the two-point function of the massless interacting scalar field, could help to make progress in this direction.  

\subsubsection*{Acknowledgement}
A.~Sanfilippo is grateful to Renato Fonseca and Patrick Hager for useful discussions. This work has been supported in part by the Excellence Cluster ORIGINS funded by the Deutsche Forschungsgemeinschaft under Grant No.~EXC - 2094 - 390783311. The work of A.~Sanfilippo was also supported by the grants
EUR2024.153549, CNS2024-154834, PID2022-139466NB-C21 and PID2025-170216NB-I00 (FEDER/UE) funded by
the Spanish Research Agency (MICIU/AEI/10.13039/501100011033).


\appendix

\section{Properties of the hyperspherical harmonics}
\label{app::hyperY}

In this Appendix we collect the relevant properties of the hyperspherical harmonics $Y_{\vec L}(x)$ using conventions appropriate for our setting. 
In $d$ dimensions they are labelled 
by the integer index vector $\vec L=(L,L_{d-1}, \ldots, L_1)$ with 
$L\geq L_{d-1} \geq \ldots \geq L_2 \geq |L_1|$.
We define the hyperspherical harmonics as an orthogonal,
\begin{equation}
\int\der^dx\;\sqrt{g(x)}Y_{\vec L}(x)Y^*_{\vec L'}(x)=\frac{1}{H^d}\delta_{\vec L\vec L'}\,,
\label{eq::Yortho}
\end{equation}
and complete,
\begin{equation}
\sum_{\vec L}Y_{\vec L}(x)Y^*_{\vec L}(y)=\frac{1}{H^d\sqrt{g(x)}}\delta^{(d)}(x-y)\,,
\label{eq::Ycomp}
\end{equation}
set of functions on the sphere, which satisfy the eigenvalue equation
\begin{equation}
\Box_xY_{\vec L}(x)=-H^2L(L+d-1)Y_{\vec L}(x)\,.
\label{eq::Yeigen}
\end{equation}
The lowest harmonic $Y_{\vec 0}$ is a constant, with squared absolute value
\begin{equation}
|Y_{\vec 0}|^2=\frac{1}{V_dH^d}
\label{eq::Y0}
\end{equation}
as a consequence of \eqref{eq::Yortho}. Combining this result with \eqref{eq::Yortho} implies the equation
\begin{equation}
\int\der^dx\;\sqrt{g(x)}\,Y_{\vec L}(x)=0\,,\quad \vec L\neq0\,.
\label{eq::Yvanish}
\end{equation}

\section{Leading large-$|Z|$ behaviour of chain integrals}
\label{app::chainz}

In this Appendix we give the details for the derivation of the leading-logarithmic behaviour of the chain integrals presented in the main text.

We start from the defining equations
\begin{flalign}
I_{0\textrm{-chain}}(Z)&\equiv\hat G(x,y)\,,\\
I_{n\textrm{-chain}}(Z)&\equiv\int_{z_1,...,z_n}\hat G(x,z_1)\times...\times \hat G(z_n,y)\,,\quad n\in\N\,.
\label{eq::appchain}
\end{flalign}
Using the representation \eqref{eq::Gspec} of $\hat G(x,y)$ and the orthogonality \eqref{eq::Yortho} of the hyperspherical harmonics,  we can reduce the $n$ integrals appearing in \eqref{eq::appchain} to the following sum:
\begin{equation}
I_{n\textrm{-chain}}(Z)=H^{d-2}\sum_{\vec L\neq0}\frac{Y_{\vec L}(x)Y^*_{\vec L}(y)}{L^{n+1}(L+d-1)^{n+1}}\,.
\end{equation}
By introducing the auxiliary, dimensionless parameter $\tilde m$, this sum can be rewritten as
 \begin{equation}
I_{n\textrm{-chain}}(Z)=\frac{(-1)^n}{n!}\frac{\p^n}{\p\tilde m^n}\bigg|_{\tilde m=0}H^{d-2}\sum_{\vec L\neq0}\frac{Y_{\vec L}(x)Y^*_{\vec L}(y)}{L(L+d-1)+\tilde m}\,.
\end{equation}
The above sum times $H^{d-2}$ yields the free non-zero mode two-point function with mass $m^2=H^2\tilde m$, which we call $\hat G_{\tilde m}(x,y)$. Using the result for the free massive Euclidean two-point function $G_{\tilde m}(x,y)$ \cite{Marolf:2010zp},
\begin{equation}
G_{\tilde m}(x,y)\equiv\frac{H^{d-2}}{(4\pi)^{\frac{d}{2}}}\frac{\Gamma(\frac{d-1}{2}+\tilde\nu)\Gamma(\frac{d-1}{2}-\tilde\nu)}{\Gamma(\frac{d}{2})}\,{_2F_1}\bigg(\frac{d-1}{2}+\tilde\nu,\frac{d-1}{2}-\tilde\nu;\frac{d}{2};\frac{1+Z}{2}\bigg)\,,
\label{eq::mprop}
\end{equation}
where $_2F_1(a,b;c;z)$ denotes Gauss' hypergeometric function, and
\begin{equation}
\tilde\nu\equiv\sqrt{\bigg(\frac{d-1}{2}\bigg)^{\!2}-\tilde m}\,,
\end{equation}
we find
\begin{equation}
\hat G_{\tilde m}(x,y)=G_{\tilde m}(x,y)-\frac{H^{d-2}}{\tilde m}\frac{\Gamma(\frac{d+1}{2})}{2\pi^{\frac{d+1}{2}}}\,.
\label{eq::tildeGhat}
\end{equation}
Any chain integral can therefore be evaluated by taking derivatives of \eqref{eq::tildeGhat}, and then setting $\tilde m=0$. It follows that the chain integral $I_{n\textrm{-chain}}(x,y)$ is related to the coefficient of the $n$-th term in the Taylor expansion of $\hat G_{\tilde m}(x,y)$ around $\tilde m=0$ as
\begin{equation}
\hat G_{\tilde m}(x,y)=\sum_{n=0}^{\infty}(-1)^nI_{n\textrm{-chain}}(Z)\,\tilde m^n\,.
\label{eq::TaylorhatG}
\end{equation}
This equation holds also if we analytically continue it in $Z$. The leading behaviour for $|Z|\rightarrow\infty$ of the free massive propagator \eqref{eq::mprop} for $\tilde m\ll1$ reads
\begin{equation}
\lim\limits_{|Z|\rightarrow\infty}G_{\tilde m}(x,y)=\frac{1}{V_dH^2\tilde m}\bigg(\!-\frac{Z}{2}\bigg)^{\!-\frac{\tilde m}{d-1}}\,.
\label{eq::secmass}
\end{equation}
Expanding \eqref{eq::secmass} for small $\tilde m$ yields, at leading-logarithmic order,
\begin{equation}
\lim\limits_{|Z|\rightarrow\infty}G_{\tilde m}(x,y)=\frac{1}{V_dH^2}\sum_{n=0}^{\infty}\frac{(-1)^n\tilde m^{n-1}}{n!}\bigg(\frac{1}{d-1}\bigg)^{\!n}\log^{n}|Z|\,.
\label{divseries}
\end{equation}
Comparing this expression to \eqref{eq::TaylorhatG} we find the leading large-$|Z|$ behaviour for the chain integrals:
\begin{equation}
\lim\limits_{|Z|\rightarrow\infty}I_{n\textrm{-chain}}(Z)=-\frac{1}{H^{2n+2}V_d(n+1)!}\bigg(\frac{1}{d-1}\bigg)^{n+1}\log^{n+1}|Z|\,,\quad n\in\N_0\,.
\end{equation}

\section{Large-order behaviour of the $\sG_n$}
\label{app::Gn}

In this Appendix we derive the estimate for the leading large-order behaviour of $\sG_n$ given in \eqref{eq:Gnasy} in the main text. 

The formula \eqref{eq::Gdimless} for $\sG_n$, after evaluating the $\tilde\vp$-integral, yields a sum of the form
\begin{equation}
\sG_n=\sum_{l=0}^{n+1}a_l(n)H^{-\frac{ld}{2}}\lambda^{\frac{l}{2}}\cor{}{\bar\phi^{2l}}\,.
\label{eq::gnansatz}
\end{equation}
The stochastic one-point functions are known in closed form from \eqref{eq::eqPDF}:
\begin{equation}
\cor{}{\bar\phi^{2n}}=\bigg(\frac{4!}{\lambda V_d}\bigg)^{\frac{n}{2}}\frac{\Gamma(\frac{2n+1}{4})}{\Gamma(\frac{1}{4})}\,.
\label{eq::stocheq}
\end{equation}
This fixes the $n$-dependence of the $\cor{}{\bar\phi^{2n}}$, so we have to determine how the coefficients $a_l(n)$ in \eqref{eq::gnansatz} depend on $n$. To this end we consider how the differential operator $\hat\Gamma_{\tilde\vp}$, defined in \eqref{eq::stochdimless}, acts on $\tilde\vp^n$:
\begin{equation}
\hat\Gamma_{\tilde\vp}\tilde\vp^n=\frac{1}{d-1}\bigg[-\frac{n}{6}\tilde\vp^{n+2}+\frac{n(n-1)}{V_dH^d}\tilde\vp^{n-2}\bigg]\,.
\end{equation}
The operator $\hat\Gamma_{\tilde\vp}$ generates two terms containing powers of $\tilde\vp$ increased and decreased by two units, respectively, where the coefficient of the lower power of $\tilde\vp$ contains one more power of $n$. This additional power of $n$ is compensated by
\begin{equation}
\frac{\cor{}{\bar\phi^{n+4}}}{\cor{}{\bar\phi^n}}\sim n\,,
\end{equation}
see \eqref{eq::stocheq}. Therefore, the quantity
\begin{equation}
\int_{-\infty}^{\infty}\der\tilde\vp\;\tilde\rho_{1\textrm{eq}}(\tilde\vp)\tilde\vp\hat\Gamma_{\tilde\vp}\tilde\vp^{n-1}\sim n\cor{}{\bar\phi^{n+2}}\sim n^2\cor{}{\bar\phi^{n-2}}
\end{equation}
scales homogeneously with $n$. From this observation, we may conclude that the combination $\tilde\vp(\hat\Gamma_{\tilde\vp})^n\tilde\vp$ appearing in the integrand of \eqref{eq::stochdimless} will give rise to a polynomial in $\tilde\vp$ which, after being weighted with $\tilde\rho_{1\textrm{eq}}(\tilde\vp)$ and integrated, will also scale homogeneously with $n$. To determine how $\sG_n$ scales with $n$, we can thus select any one of the terms in \eqref{eq::gnansatz} and determine its $n$-scaling. We choose the terms which are generated solely by the single-derivative terms:
\begin{equation}
\sG_n\sim\frac{1}{n!}\int_{-\infty}^{\infty}\der\tilde\vp\;\tilde\vp\tilde\rho_{1\textrm{eq}}(\tilde\vp)\bigg[\tilde\vp^3\frac{\p}{\p\tilde\vp}\bigg]^n\tilde\vp=\frac{(2n-1)!!}{n!}\bigg(\frac{4!}{V_dH^d}\bigg)^{\!n+1}\frac{\Gamma(\frac{2n+3}{4})}{\Gamma(\frac{1}{4})}\,.
\end{equation}
This fixes the leading $n$-scaling of $\sG_n$, up to an overall real factor $A$ and the $n$-th power of a different real constant $C$. Thus, we can approximate for large values of $n$
\begin{equation}
\lim\limits_{n\rightarrow\infty}\sG_n\approx A\,C^n\frac{(2n-1)!!}{n!}\,\Gamma\bigg(\frac{2n+3}{4}\bigg)\,,\quad A\,,C\in\R\,.
\end{equation}

\section{Analytic formulae for $A$ and $C$}
\label{app:ACderivation}

In this Appendix we give the details of the derivation of the analytic formulae for the constants $A$ and $C$ given in the main text. 

Defining 
\begin{equation}
\tilde\vp\equiv\bigg(\frac{\lambda}{H^{\frac{d}{2}}}\bigg)^{\frac{1}{4}}\vp\,,\quad\lambda_n\equiv\frac{\Lambda_n}{H^{\frac{d}{2}-1}\sqrt{\lambda}}\,,
\end{equation}
we rewrite \eqref{eq::Schrodi} in terms of dimensionless eigenfunctions and -values as
\begin{equation}
\bigg[-\frac{\p^2}{\p\tilde\vp^2}+\underbrace{\frac{(V_dH^d)^2}{144}\tilde\vp^6-\frac{V_dH^d}{4}\tilde\vp^2}_{\equiv U(\tilde\vp)}\bigg]\Phi_n(\tilde\vp)=(d-1)V_dH^d\lambda_n\Phi_n(\tilde\vp)\,.
\end{equation}
In the limit $n\to\infty$ the WKB approximation to both the $\Phi_n(x)$ and the $\lambda_n$ becomes exact. Therefore, we may compute the leading large-$n$ form of the $\lambda_n$ using the WKB formula \cite{Bender:1999box}
\begin{equation}
\int_{-\tilde\vp_{0,n}}^{\tilde\vp_{0,n}}\der \tilde\vp\;\sqrt{(d-1)V_dH^d\lambda_n-U(\tilde\vp)}=\pi\bigg(n+\frac{1}{2}\bigg)\,,
\label{eq:WKBlambda}
\end{equation}
where $\pm\tilde\vp_{0,n}$ denotes the two turning points determined by 
\begin{equation}
U(\pm \tilde\vp_{0,n})-(d-1)V_dH^d\lambda_n=0\,.
\end{equation}
In the large-$n$ limit the sextic term in $U(\tilde\vp)$ dominates, and the integral \eqref{eq:WKBlambda} can be evaluated exactly. It yields the leading asymptotic behaviour
\begin{equation}
\lim\limits_{n\to\infty}\lambda_n=\frac{16\pi^3}{9\Gamma(\frac{1}{3})^{\frac{9}{2}}(d-1)\sqrt{V_dH^d}}\,n^{\frac{3}{2}}\,.
\label{eq:leadingln}
\end{equation}

Similarly, the leading large-$n$ form of the dimensionless coefficients
\begin{equation}
\beta_n\equiv\frac{\sqrt{\lambda}}{H^{\frac{d}{2}}}\,b_n=\bigg[\int_{-\infty}^{\infty}\der\tilde\vp\;\tilde\vp\,\Phi_0(\tilde\vp)\Phi_n(\tilde\vp)\bigg]^2\,,
\label{eq:WKBbn}
\end{equation}
where we used \eqref{eq::bn}, may be computed using the WKB-approximate form for $\Phi_n(\tilde\vp)$ with $n>0$. However, this requires a detailed analysis covering the contributions of the oscillating, turning-point, and exponentially-suppressed regions of the WKB approximation of $\Phi_n(\tilde\vp)$ to the integral in \eqref{eq:WKBbn}. 
With the help of Claude Opus 5.0 and Opus 5.5 we found the following analytic formula for the leading asymptotic behaviour of the $\beta_n$:
\begin{equation}
\lim\limits_{n\to\infty}\beta_n=2^{\frac{1}{4}}\frac{16\cdot 3^{\frac{3}{4}}\pi^{\frac{3}{2}}}{\Gamma(\frac{1}{4})\Gamma(\frac{1}{3})^{\frac{9}{4}}\sqrt{V_dH^d}}\,n^{-\frac{1}{4}}\exp\bigg(-\frac{\pi}{\sqrt{3}}n\bigg)\,.
\label{eq:leadingbn}
\end{equation}

Given \eqref{eq:leadingln} and \eqref{eq:leadingbn}, the leading large-$n$ asymptotics of the $\sG_n$ may be computed from
\begin{equation}
\sG_n=\frac{1}{n!}\sum_{\textrm{odd }m>0}\beta_m(-\lambda_m)^n\,.
\end{equation}
For $n\to\infty$ this sum is dominated by large values of $m$. For such large values, each summand scales like
\begin{equation}
(-1)^n\exp\!\left(-\frac{\pi}{\sqrt{3}}m\right)m^{\frac{3n}{2}-\frac{1}{4}}\,.
\end{equation}
We can use this scaling for all summands, since the error is a subleading effect in the $n\to\infty$ limit.
For such large values of $m$ the spacing between the integers becomes negligible with respect to the size of the integers themselves, and the sum is well-approximated by the integral
\begin{flalign}
\sum_{\textrm{odd }m>0}m^{\frac{3n}{2}-\frac{1}{4}}\exp\bigg(-\frac{\pi}{\sqrt{3}}m\bigg)&\approx\frac{1}{2}\int_0^{\infty}\der x\;x^{\frac{3n}{2}-\frac{1}{4}}\exp\bigg(-\frac{\pi}{\sqrt{3}}x\bigg)\nonumber\\
&=\frac{1}{2}\,\bigg(\frac{\pi}{\sqrt{3}}\bigg)^{\!-\frac{3n}{2}-\frac{3}{4}}\,\Gamma\bigg(\frac{3n}{2}+\frac{3}{4}\bigg)\,,
\end{flalign}
where the factor of $\frac{1}{2}$ in front of the integral accounts for the fact that the original sum runs only over odd (positive) integers. Applying the Stirling approximation 
\begin{equation}
\lim\limits_{n\to\infty}\,\frac{1}{n!}\,\Gamma\bigg(\frac{3n}{2}+\frac{3}{4}\bigg)=\bigg(\frac{3}{2}\bigg)^{\!\frac{3n}{2}+\frac{1}{4}}e^{-\frac{n}{2}}\,n^{\frac{n}{2}-\frac{1}{4}}\,,
\end{equation}
and putting the pieces together, we find the leading asymptotic estimate
\begin{equation}
\lim\limits_{n\to\infty}\sG_n\approx2^{\frac{1}{4}}\cdot\frac{12\cdot 3^{\frac{3}{8}}(2\pi)^{\frac{3}{4}}}{\Gamma(\frac{1}{4})\Gamma(\frac{1}{3})^{\frac{9}{4}}\sqrt{V_dH^d}}\bigg[-\sqrt{2}\frac{4\cdot3^{\frac{1}{4}}\pi^{\frac{3}{2}}}{\Gamma(\frac{1}{3})^{\frac{9}{2}}(d-1)\sqrt{V_dH^d}}\,\bigg]^n e^{-\frac{n}{2}}n^{\frac{n}{2}}\,.
\end{equation}
Comparing this result with the form of \eqref{eq::estimatestandardform},
\begin{equation}
\lim\limits_{n\rightarrow\infty}\sG_n=2^{\frac{1}{4}}A\,(\sqrt{2}C)^n\,e^{-\frac{n}{2}}n^{\frac{n}{2}}\,,
\end{equation}
we read off the analytic expressions \eqref{eq:A}, \eqref{eq:C} for $A$ and $C$, respectively.

\section{Nevanlinna-Sokal theorem}
\label{app:NStheorem}

\paragraph{Theorem (Nevanlinna-Sokal \cite{Sokal:1980ey})}
Let $R>0$ and let
\begin{equation}
  C_R=\bigg\{\tau\in\C\;\bigg|\;\Re\bigg(\frac{1}{\tau}\bigg)>\frac{1}{R}\bigg\}
\end{equation}
be the open disc of diameter $R$ with one boundary point at the origin, tangent there to the
imaginary axis. Suppose $f$ is holomorphic on $C_R$ and there are $A\,,\sigma>0$ such that
\begin{equation}\label{eq:nshyp}
  \left|f(\tau)-\sum_{n=0}^{N-1}a_n\tau^n\right|\leq A\,\sigma^N\,N!\,|\tau|^N
  \qquad\text{for all }N\geq0\,,\ \tau\in C_R\,.
\end{equation}
Then:
\begin{enumerate}[label=\alph*)]
\item $B(t)=\sum_na_nt^n/n!$ converges in the circle $|t|<1/\sigma$;
\item $B(t)$ has an analytic continuation to the strip
$\mathcal{S}_{\sigma}=\{t\,|\,\textrm{dist}(t,\R_+)<1/\sigma\}$, and satisfies the bound
\begin{equation}
|B(t)|\leq K\exp\bigg(\frac{|t|}{R}\bigg)
\label{eq:Borelbound}
\end{equation}
uniformly in every $\mathcal{S}_{\sigma'}$ with $\sigma'>\sigma$;
\item For every $\tau\in C_R$ the integral 
\begin{equation}
  F(\tau)=\frac1\tau\int_0^\infty\der t\; e^{-\frac{t}{\tau}}B(t)
  \label{eq:borelsum}
\end{equation} converges absolutely and
equals $f(\tau)$.
\end{enumerate}
Conversely, if $B(t)$ is holomorphic on $\mathcal{S}_{\sigma''}$, with $\sigma''<\sigma$, and satisfies the bound given in \eqref{eq:Borelbound} for $t\in\mathcal{S}_{\sigma''}$, then
\eqref{eq:borelsum} defines a holomorphic function $F(\tau)$ on $C_R$ and satisfies \eqref{eq:nshyp} on $C_{R'}$ with $R'<R$. 
\bibliography{Refs}{}

\end{document}